\documentclass[prb, twocolumn]{revtex4-2}
\usepackage{amsfonts}
\usepackage{amsmath}
\usepackage{amssymb}
\usepackage{graphicx}
\usepackage{epstopdf}
\usepackage{color}

\input epsf.sty

\begin{document}

\title{Quantum oscillations of Kondo screening phases in strong magnetic fields}
\author{Po-Hao Chou$^{1}$, Chung-Hou Chung$^{3}$, and  Chung-Yu Mou$^{1,2,4}$}

\affiliation{$^{1}$Physics Division, National Center for Theoretical Sciences, Taipei 10617, Taiwan, R.O.C}
\affiliation{$^{3}$Electrophysics Department, National Yang Ming Chiao Tung University, Hsinchu, Taiwan 300, R.O.C.}
\affiliation{$^{2}$Center for Quantum Technology and Department of Physics, National Tsing Hua University, Hsinchu, Taiwan 300, R.O.C.}
\affiliation{$^{4}$Institute of Physics, Academia Sinica, Nankang, Taiwan, R.O.C.}

\begin{abstract}
We generalize the iterative diagonalization procedure adopted in method of numerical renormalization group to analyze the Kondo effect in strong magnetic fields, where the density of states for itinerary electrons at the chemical potential varies discontinuously as the magnetic field changes. We first examine phases of many-body ground states in the presence of single impurity. By investigating change of $z$-component of total spin, $\Delta S_z$, and spin-spin correlation between the impurity and conduction electrons, we find that there are three states competing for the ground state when Zeeman splitting is present. One of the states is doublet in which the impurity spin is unscreened. The other two states are Kondo screening states with $\Delta S_z=1/2$ and $\Delta S_z=1$, in which the impurity spin is partially screened and completely screened respectively.  For Kondo systems with two-impurities in strong magnetic fields, we find that the interplay between the Kondo screening effect, RKKY interaction, and quantum oscillations due to Landau levels determines the ground state of the system. Combination of these three factors results in different screening scenarios for different phases in which spins of two impurities can form spin-0 or spin-1 states, while impurity spins in these phases can be either screened, partially screened, or unscreened by conduction electrons. The emergence of the ground state from these competing states oscillates with the change of magnetic field, chemical potential or inter-impurity distance. This leads to quantum oscillations in magnetization and conductivity.  In particular, we find extra peak structures in longitudinal conductivity that reflect changes of Kondo screening phases and are important features to be observed in experiments.  Our results provide a complete characterization of phases for Kondo effect in strong magnetic fields.
\end{abstract}
\maketitle
\section{Introduction}
How the magnetic order emerges from the interaction between localized magnetic moments and itinerary electrons is an important issue to understand magnetism in correlated metals. The issue has been clarified at the level of single magnetic moment, in which the moment gets screened and it results in a correlated Kondo  screening state\cite{Hewson,mou1}. It is further realized that when number of magnetic moments exceeds one, the Rudermann–Kittel–Kasuya–Yoshida (RKKY) interaction is induced between moments, which starts to compete with the Kondo effect.  Depending on the distance between moments, the induced coupling between two magnetic moments oscillates between ferromagnetic (FM) or antiferromagnetic (AFM) coupling, leading to complicated competitions among correlated singlet state, triplet state and Kondo screening states\cite{Doniach, Coqblin, twoimpurity1, twoimpurity2, twoimpurity3,twoimpurity4, twoimpurity5, twoimpurity6}.

While the above understandings have been known for a while, they were based on the analysis in conventional metallic systems in which
 the density of states (DOS) for itinerary electrons is nearly a constant near the chemical potential $\mu$.  When external magnetic fields are in presence, the assumption of constant DOS breaks down. In the case when magnetic fields are weak, the degeneracy at the chemical potential for different spin components is lifted. This leads to the differentiation of possible correlated states for two impurities and makes these correlated states observable in experiments\cite{ImpMag1}. More recently, the de Haas-Van Alphen effect is observed  for Kondo insulators in strong magnetic fields\cite{Li, Sebastian,Sebastian2}.  The oscillation is shown to result from the emergence of Landau levels in electronic structures\cite{mou2}.  In this case, the density of states for itinerary electrons at the chemical potential varies discontinuously as the magnetic field changes. Consequently, the Kondo screening effect should be entirely different.  For instance, when $\mu$ lies at the middle between two Landau levels without particle-hole symmetry, as suggested by Kondo effects studied for gapped systems (such as semiconductors)\cite{ImpGap1,ImpGap2,ImpGap3}, one expects that the system should undergo a quantum phase transition from singlet to doublet if the spacing between Landau levels is compared with the Kondo temperature, leading to the breakdown of screening effect at low temperature. Here the energy gap in the gapped system plays a similar role as the spacing of Landau levels for Kondo systems in strong magnetic fields. In addition, the Kondo effect is shown to exhibit re-entrant behavior as the chemical potential changes\cite{LLSlaveB}. Perturbative studies of two Kondo impurities in graphene indicate that generic competition between Kondo screening and the RKKY interaction persists even with Landau levels being in presence\cite{RKKY1,RKKY2,RKKY3, LLRKKY, LLSlaveB,Bulla, LLRKKY}. Nonetheless, the complete phase diagram and behaviors of relevant physical quantities such like entropy, specific heat and susceptibility are still unknown.

Theoretically, in the absence of Landau levels, the numerical renormalization group (NRG) method has provided more complete description of the Kondo screening than the mean-field and perturbation approach\cite{Bulla}. In this paper, we borrow the iterative diagonalization procedure from NRG method to investigate one and two magnetic
impurities screened by discrete Landau levels at zero temperature.
We will show that the ground state generally oscillates in Kondo screening state, partially-screened, and unscreened spin states. This leads to quantum oscillations observed in magnetization of the system. In particular, we find that two-impurities in the ground state can form spin 0 (singlet) and spin 1 (triplet) states. Remarkably,
these states can be either screened, partially screened, or unscreened with the emergence of these states being oscillating with the change of magnetic field, chemical potential or inter-impurity distance.  Our results indicate that the oscillation in Kondo screening phases is the key to understand the observed quantum oscillation in Kondo systems.

\section{Model Hamiltonians}
We start by considering the two dimensional multi-impurities Anderson model with the magnetic field $\vec{B}$ being along z-direction. By treating the conduction electron in the continuum limit, the Hamiltonian can be written as
\begin{eqnarray} \label{H0} \notag
&& H  = H_c + H_d + H_V, \\ \notag
&& H_c = \int d\vec{r} \sum_{\sigma}c_{\vec{r}\sigma}^{\dag}(\frac{\vec{\Pi}^2}{2m^*_e}-\mu)c_{\vec{r}\sigma}+g_c\mu_BB s^{z}_{c,\vec{r}}, \\ \notag
&& H_d = U\sum_{j} n^d_{j\uparrow}n^d_{j\downarrow}+\sum_{j\sigma}\xi^d n^d_{j\sigma}+g_d\mu_BB\sum_{j} s^{z}_{d,j}, \\
&& H_V = Va \int d\vec{r} \sum_{j\sigma} \delta(\vec{r}-\vec{r}^{\:d}_j) (d_{j\sigma}^{\dag}c_{\vec{r}\sigma}+h.c.).
\end{eqnarray}
Here $H_c$ is the Hamiltonian for describing the conduction electrons, $H_d$ is the Hamiltonian for describing the impurities' electrons, and $H_V$ is describing the hybridization between conduction and impurity's electrons.
$\vec{\Pi}=\vec{p}+e\vec{A}/c$ is the kinetic momentum operator with $\vec{A}$ being the vector potential for $\vec{B}$,
$\mu$ is the chemical potential, $m^*_e$ is the effective mass of the electron, $a^2$ is the effective area of impurity hybridization range, $g_c$ and $g_d$ are the g-factors of
conduction and impurities' electrons, and $V$ is the hybridization strength between impurities and conduction electrons.
$c_{\vec{r}\sigma}^{\dag}$ and $d_{j\sigma}^{\dag}$ are the creation operators for conduction electron at position $\vec{r}=(x_j^d,y_j^d)$ and localized electron at position $\vec{r}^{\:d}_j$ with spin $\sigma$ respectively.
$s^{z}_{c,\vec{r}}=\frac{1}{2}(c_{\vec{r}\uparrow}^{\dag}c_{\vec{r}\uparrow}-c_{\vec{r}\downarrow}^{\dag}c_{\vec{r}\downarrow})$ and
$s^{z}_{d,j}=\frac{1}{2}(d_{j\uparrow}^{\dag}d_{j\uparrow}-d_{j\downarrow}^{\dag}d_{j\downarrow})$ are spin operators for conduction and impurities' electrons respectively.
In the Landau gauge $\vec{A}=(0,Bx,0)$, the single particle eigen-energy and the corresponding eigenfunction of the conduction electrons are
\begin{eqnarray} \notag
&&\varepsilon^c_n=\varepsilon_B(n+\frac{1}{2}),\:\:\xi^c_n=\varepsilon^c_n-\mu,\\ \notag
&&\psi_{n,k_y}(\vec{r})=\frac{e^{-ik_y y}}{\sqrt{L}}\phi_n(x+x_k),x_k=l_B^2k_y, \\
&& \phi_n(x)=\frac{1}{\sqrt{2^nn!\pi^{1/2}l_B}}H_n(x/l_{B})e^{-x^2/(2l_B^2)},
\end{eqnarray}
where $\varepsilon_B=\hbar\omega_B$ is the Landau quantized energy with $\omega_B=eB/(m^*_ec)$ being the cyclotron frequency,
$k_y$ is the wave-vector along $y$ direction, $l_B=\sqrt{\hbar/(m^*_e\omega_B)}=\sqrt{c\hbar/(eB)}$ is the magnetic length,
and $H_n(x)$ is the $n_{th}$ Hermite polynomial.
The magnetic length $l_B$ is approximated to $25.7\: \mbox{nm}/ \sqrt{B(T)}$, where $B(T)$ represents the magnetic field $B$ is unit of Tesla.
The restriction of $-L/2\leq x_k\leq L/2$ gives the allowed states number of $k_y$ as Landau degeneracy
\begin{eqnarray}\label{NL0}
N_L=L^2/(2\pi l_B^2)=L^2\frac{m^*_e}{2\pi\hbar^2}\varepsilon_B=L^2\rho \varepsilon_B,
\end{eqnarray}
where $L$ is the length of square system, $\rho=m^*_e/(2\pi\hbar^2)$ is the density of states of two dimensional free electron gas with effective mass $m^*_e$.

By using $\psi_{n,k_y}(\vec{r})$, we can transform the annihilation operator to the basis of the Landau quantized states as
\begin{eqnarray} \label{C0}\notag
&& c_{nk_y\sigma}=\int d\vec{r}\: \psi_{n,k_y}(\vec{r})c_{\vec{r}\sigma}, \\
&& c_{\vec{r}\sigma}=\sum_{nk_y}\psi^{*}_{n,k_y}(\vec{r})c_{nk_y\sigma}.
\end{eqnarray}

After applying the transformation, the Hamiltonian projected in Landau eigen-states is given by
\begin{eqnarray} \label{H00}
&& H = H_d +\sum_{\{n\},k_y\sigma} \xi^c_{n\sigma} c_{nk_y\sigma}^{\dag} c_{nk_y\sigma} \\ \notag
&&+\frac{\tilde{V}}{\sqrt{L}} \sum_{j\{n\},k_y\sigma} e^{ik_yy_j^d}\phi_n(x_j^d+x_k)d_{j\sigma}^{\dag}c_{nk_y\sigma}+h.c..
\end{eqnarray}
Here  $\tilde{V}=Va$, $\xi^c_{n\uparrow}=\xi^c_{n}+g_c\mu_BB/2$, $\xi^c_{n\downarrow}=\xi^c_{n}-g_c\mu_BB/2$, and we also take an energy cutoff $D$ so that $\xi^c_n=\varepsilon^c_n-\mu \in [-D,D ]$ and $n=\{N_{min},N_{min}+1,...,N_{max}\}$ with  $N_{max}$ being the maximum number and $N_{min}$ being the minimum number of the Landau level index $n$.

\subsection{Reduction in degrees of freedom for conduction electrons coupling with impurities} \label{reduction}
Before further simplifying the Hamiltonian, we shall first show that degrees of freedom for conduction electrons coupling with impurities can be reduced. This is illustrated by considering a toy model in which an impurity couples to two degenerate one-dimensional chains, $X_n$ and $ Y_n$, with the Hamiltonian 
\begin{eqnarray} 
&&H_T = H_d + \sum_{\{n\},\sigma} \xi_{n\sigma} (X_{n\sigma}^{\dag}X_{n\sigma}+Y_{n\sigma}^{\dag}Y_{n\sigma}) \nonumber \\ 
&& +\sum_{\{n\},\sigma} d_{\sigma}^{\dag}(V_X X_{n\sigma}+V_Y Y_{n\sigma})+h.c.. 
\end{eqnarray}
By defining two new operators 
\begin{eqnarray} \notag
&& A_{n\sigma}=\frac{1}{\sqrt{V_X^2+V_Y^2}}(V_X X_{n\sigma}+V_Y Y_{n\sigma}), \nonumber \\ 
&& B_{n\sigma}=\frac{1}{\sqrt{V_X^2+V_Y^2}}(-V_Y X_{n\sigma}+ V_X Y_{n\sigma}),
\end{eqnarray}
it is then easy to see that $A_{n\sigma}$ and $B_{n\sigma}$ obey Fermionic commutation relations: $\{A^{\dag}_{n\sigma},A_{n\sigma}\}=\{B^{\dag}_{n\sigma},B_{n\sigma}\}=1,\: \{A^{\dag}_{n\sigma},B_{n\sigma}\}=\{A_{n\sigma},B_{n\sigma}\}=0$. The Hamiltonian $H_T$ can be re-written as
\begin{eqnarray} 
&&H_T= H_d + \sum_{\{n\},\sigma} \xi_{n\sigma} (A_{n\sigma}^{\dag}A_{n\sigma}+B_{n\sigma}^{\dag}B_{n\sigma}) \notag \\
&&+\sqrt{V_X^2+V_Y^2}\sum_{\{n\},\sigma} d_{\sigma}^{\dag}A_{n\sigma}+h.c..
\end{eqnarray}
Clearly, we see that the operator $B_{n\sigma}$ decouples from the impurity and only $A_{n\sigma}$ couples to the impurity with a stronger hybridization strength $\sqrt{V_X^2+V_Y^2}$. Effectively, degrees of freedom for conduction electrons coupling with impurities is reduced. 

Going back to the Hamiltonian, Eq.(\ref{H00}), the impurity operator $ d^{\dag}_{j\sigma}$ also couples to two degenerated states operators $X_{n}$ and $Y_{n}$ with a $2$-dimensional hybridization vector $\vec{V}=(V_X,V_Y)$. Hence similar reduction of degrees of freedom can be performed (see the following subsection). In general, if the conduction electrons possess more degeneracies characterized by $N_L$ ($N_L=2$ for $H_T$, for Landau levels, $N_L$ is the Landau degeneracy given by Eq. (\ref{NL0})), the coupling of conduction electrons to a single impurity can be characterized by a $N_L$- dimensional hybridization vector $\vec{V}$. By performing similar analysis, it is clear that an impurity effectively only couples to one channel with hybridization strength $\|\vec{V}\|$. Furthermore, if there are $N_{imp}$ impurities operators  $d^{\dag}_{j}$  coupling to $N_L$ degenerated conduction electrons with the hybridization vector $\vec{V}_j$, these impurities effectively couple to $N_L-N_{imp}$ channel when $N_L\geq N_{imp}$.

\subsection{Reduced single impurity Hamiltonian $H_1$}
We start with the single impurity case with the position of the impurity being at $(0,R)$.
The hybridization between the impurity and conduction electrons is given by
\begin{eqnarray}
H_V= \frac{\tilde{V}}{\sqrt{L}} \sum_{\{n\},k_y\sigma} \phi_n(x_k)e^{ik_yR}d_{\sigma}^{\dag}c_{nk_y\sigma}+h.c..
\end{eqnarray}
By collecting annihilation operators which couple to the impurity and redefining them as a new operator as
\begin{eqnarray}
x_nA_{n\sigma} = \sqrt{\frac{L}{N_L}}\sum_{k_y} e^{ik_y R}\phi_n(x_k)c_{nk_y \sigma},
\end{eqnarray}
where $x_n=\sqrt{\frac{L}{N_L}\sum_{k_y}|e^{ik_yR}\phi_n(x_k)|^2}$ is the normalization constant.
The new hybridization term becomes $\tilde{V}\sqrt{\rho \varepsilon_B}\sum_{\{n\} \sigma} \left(x_nd^\dag_{\sigma}A_{n\sigma}+h.c.\right)$. According to the analysis in  Sec.(\ref{reduction}), it is clear that the impurity only hybridizes to
$A_{n\sigma}$ and decouples from the remaining $N_L-1$ states if $N_L\geq1$.

To obtain the normalized constant $x_n$, we first note that the dimensionless function,
\begin{eqnarray} 
\bar{\phi}_n(x/l_B)=\sqrt{l_B}\phi_n(x)=\frac{1}{\sqrt{2^nn!\pi^{1/2}}}H_n(x/l_{B})e^{-x^2/(2l_B^2)}, \nonumber \\
\end{eqnarray}
satisfies the normalization condition $\int^{\infty}_{-\infty} dt \bar{\phi}^2_n(t)=1$. For $L\gg l_B$, the normalized constant $x_n$ can be simplified as follows
\begin{eqnarray} \notag
&&x_n=\sqrt{\frac{L}{N_L}\sum_{k_y}\phi^2_n(x_k)}=\sqrt{\frac{L^2}{2\pi N_L}\int^{k^{max}_y}_{k^{min}_y}dk_y \phi^2_n(x_k)}\\
&&=\sqrt{\int^{\frac{L}{2}}_{-\frac{L}{2}}dx_k \phi^2_n(x_k)}=\sqrt{\int^{\infty}_{-\infty}dt \bar{\phi}^2_n(t)}=1,
\end{eqnarray}
where $t=x/l_B$. Thus, we obtain the hybridization term in the new basis as $(\rho \tilde{V}^2 \varepsilon_B)^{1/2}\sum_{\{n\} \sigma} \left(d^\dag_{\sigma}A_{n\sigma}+h.c.\right)$. As a result,  the reduced single impurity Hamiltonian $H_1$ in terms of $A_{n,\sigma}^{\dag}$ is given by
\begin{eqnarray} \label{H001} \notag
H_{1} = &&Ud_{\uparrow}^{\dag}d_{\uparrow}d_{\downarrow}^{\dag}d_{\downarrow}+\sum_{\sigma} \xi^d_{\sigma} d_{\sigma}^{\dag}d_{\sigma}+
\sum^{}_{\{n\},\sigma} \xi^c_{n\sigma} A_{n\sigma}^{\dag} A_{n\sigma} \\
&&+(\frac{\Gamma\varepsilon_B}{\pi})^{1/2} (\sum_{\{n\},\sigma}d_{\sigma}^{\dag}A_{n\sigma}+h.c. ),
\end{eqnarray}
where $\Gamma=\pi\rho V^2$. It is important to note that only one channel in the Landau level couples to the impurity in $H_1$.

\subsection{Reduced two impurities Hamiltonian $H_2$}

In two impurities case, positions of impurities are set at $(0,\pm R/2)$. For general positions,  please see the Supplementary Information\cite{sup}.
The hybridization between impurities and conduction electrons is given by
\begin{eqnarray}
\frac{\tilde{V}}{\sqrt{L}} \sum_{\{n\},k_y\sigma} \phi_n(x_k)(e^{ik_yR/2}d_{1\sigma}^{\dag}+e^{-ik_yR/2}d_{2\sigma}^{\dag})c_{nk_y\sigma}+h.c.. \nonumber \\
\end{eqnarray}
Hence for impurity 1, the impurity operator $d^\dag_{1\sigma}$ couples to $\sqrt{L/N_L}\sum_{k_y} e^{ik_y R/2}\phi_n(x_k)c_{nk_y \sigma}$; while for impurity 2, the impurity operator $d^\dag_{2\sigma}$ couples to $\sqrt{L/N_L}\sum_{k_y} e^{-ik_y R/2}\phi_n(x_k)c_{nk_y \sigma}$. It is easy to see that the overlap of the coefficients in the above operators is non-vanishing
\begin{eqnarray} 
&&\int^{\infty}_{-\infty} dt \left(e^{-it (R/2l_B)}\bar{\phi}_n(t)\right)^*e^{it(R/2l_B)}\bar{\phi}_n(t) \nonumber \\ 
&&=\int^{\infty}_{-\infty} dt e^{i\eta t}\bar{\phi}^2_n(t)\neq 0,
\end{eqnarray}
where $\eta=R/l_B$. Hence these operators are not orthogonal when $L\gg l_B$. However, if we define
\begin{eqnarray} \notag
&&x_nA_{n\sigma}=\sqrt{\frac{L}{N_L}}\sum_{k_y} \cos(k_yR/2)\phi_n(x_k)c_{nk_y\sigma}, \\
&&y_nB_{n\sigma}=\sqrt{\frac{L}{N_L}}\sum_{k_y} i\sin(k_yR/2)\phi_n(x_k)c_{nk_y\sigma},
\end{eqnarray}
where $x_n$ and $y_n$ are normalization constants to be determined, we shall see that by fixing $x_n$ and $y_n$
correctly, $A_{n\sigma}$ and $B_{n\sigma}$ are two orthogonal annihilation operators with standard  Fermion commutation relations. First, note that because the following integral vanishes
\begin{eqnarray} 
\int^{\infty}_{-\infty} dt \sin (\frac{\eta t}{2})\cos (\frac{\eta t}{2}) \bar{\phi}^2_n(t)
=\frac{1}{2}\int^{\infty}_{-\infty} dt \sin (\eta t) \bar{\phi}^2_n(t)=0, \nonumber \\
\end{eqnarray}
we have $\{A^{\dag}_{n\sigma},B_{n\sigma}\}=0$ when $L\gg l_B$. The normalization constants $x_n$ and $y_n$ are determined by the required commutation relations and are given by\cite{sup}
\begin{eqnarray} 
&&x_n=\sqrt{ \frac{1}{2}\int^{\infty}_{-\infty}dt\cos^2\frac{\eta t}{2}\bar{\phi}^2_n(t)}=\sqrt{\frac{1+u_n}{2}}, \nonumber \\ 
&& y_n=\sqrt{ \frac{1}{2}\int^{\infty}_{-\infty}dt\sin^2\frac{\eta t}{2}\bar{\phi}^2_n(t)}=\sqrt{\frac{1-u_n}{2}}, \nonumber \\
&& u_n=\mathfrak{Re}\left(\int^{\infty}_{-\infty}dt e^{i\eta t}\bar{\phi}^2_n(t)\right)=e^{-\eta^2/4}L_n(\eta^2/2),
\end{eqnarray}
where $L_n(x)$ is the $n_{th}$ Laguerre polynomials. Thus the effective hybridization becomes
\begin{eqnarray}\label{hyb2} \notag
&&(\rho \tilde{V}^2 \varepsilon_B)^{1/2} \times \\ \notag
&&\left[ d^\dag_{1\sigma}(x_nA_{n\sigma}+y_nB_{n\sigma})+d^\dag_{2\sigma}(x_nA_{n\sigma}-y_nB_{n\sigma})+h.c.\right]. \\
\end{eqnarray}
The reduced two impurities Hamiltonian $H_2$ in terms of $A_{n,\sigma}^{\dag}$ and $B_{n,\sigma}^{\dag}$ is then given by
\begin{eqnarray} \label{H002} \notag
H_{2} = &&U\sum_{j=1,2}d_{j\uparrow}^{\dag}d_{j\uparrow}d_{j\downarrow}^{\dag}d_{j\downarrow}+\sum_{j=1,2,\sigma} \xi^d_{\sigma} d_{j\sigma}^{\dag}d_{j\sigma}\\ \notag
 &&+\sum^{}_{\{n\},\sigma} \xi^c_{n\sigma} \left(A_{n\sigma}^{\dag} A_{n\sigma}+B_{n\sigma}^{\dag} B_{n\sigma}\right) \\ \notag
 &&+ (\frac{\Gamma\varepsilon_B}{\pi})^{1/2}\sum_{\{n\},\sigma} [ (d_{1\sigma}^{\dag}+d_{2\sigma}^{\dag})x_nA_{n\sigma} \\
 &&\:\:\:\:\:\:\:\:\:\:\:\:\:\:\:\:\:+(d_{1\sigma}^{\dag}-d_{2\sigma}^{\dag})y_nB_{n\sigma} +h.c.].
\end{eqnarray}
Note that if we redefine  new operators $X_{n\sigma}$ and $Y_{n\sigma}$ by
performing an orthogonal transformation between $A_{n\sigma}$ and $B_{n\sigma}$ as follows
\begin{eqnarray} 
X_n=x_nA_{n\sigma}+y_nB_{n\sigma},\: Y_n=-y_nA_{n\sigma}+x_nB_{n\sigma},
\end{eqnarray}
the hybridization show in Eq.(\ref{hyb2}) becomes
\begin{eqnarray} \label{hybridization}
&&(\rho \tilde{V}^2 \varepsilon_B)^{1/2} \times \nonumber \\ 
&&\left(d^\dag_{1\sigma}X_{n\sigma}+d^\dag_{2\sigma}(u_nX_{n\sigma}-\sqrt{1-u_n^2}Y_{n\sigma})+h.c.\right).
\end{eqnarray}
It is clear from the above form of hybridization that when all $u_n$ vanish, two impurities decouples and the system is the same as the case for single impurity; while if all $u_n$ are equal to $1$, both the impurity operators couple to the same operator $X_{n\sigma}$, the system thus becomes the well-known two-impurities-one-channel Kondo problem.
In Fig.~1,  we show how $u_n$ depends on the dimensionless inter-impurity distance $\eta$. Clearly, neither all $u_n$ vanish nor $u_n$ are all equal to $1$. Hence two-impurities in the Anderson model is a two-channel problem.
\begin{figure}[tp] 
\includegraphics[height=2.2in, width=3.3in]{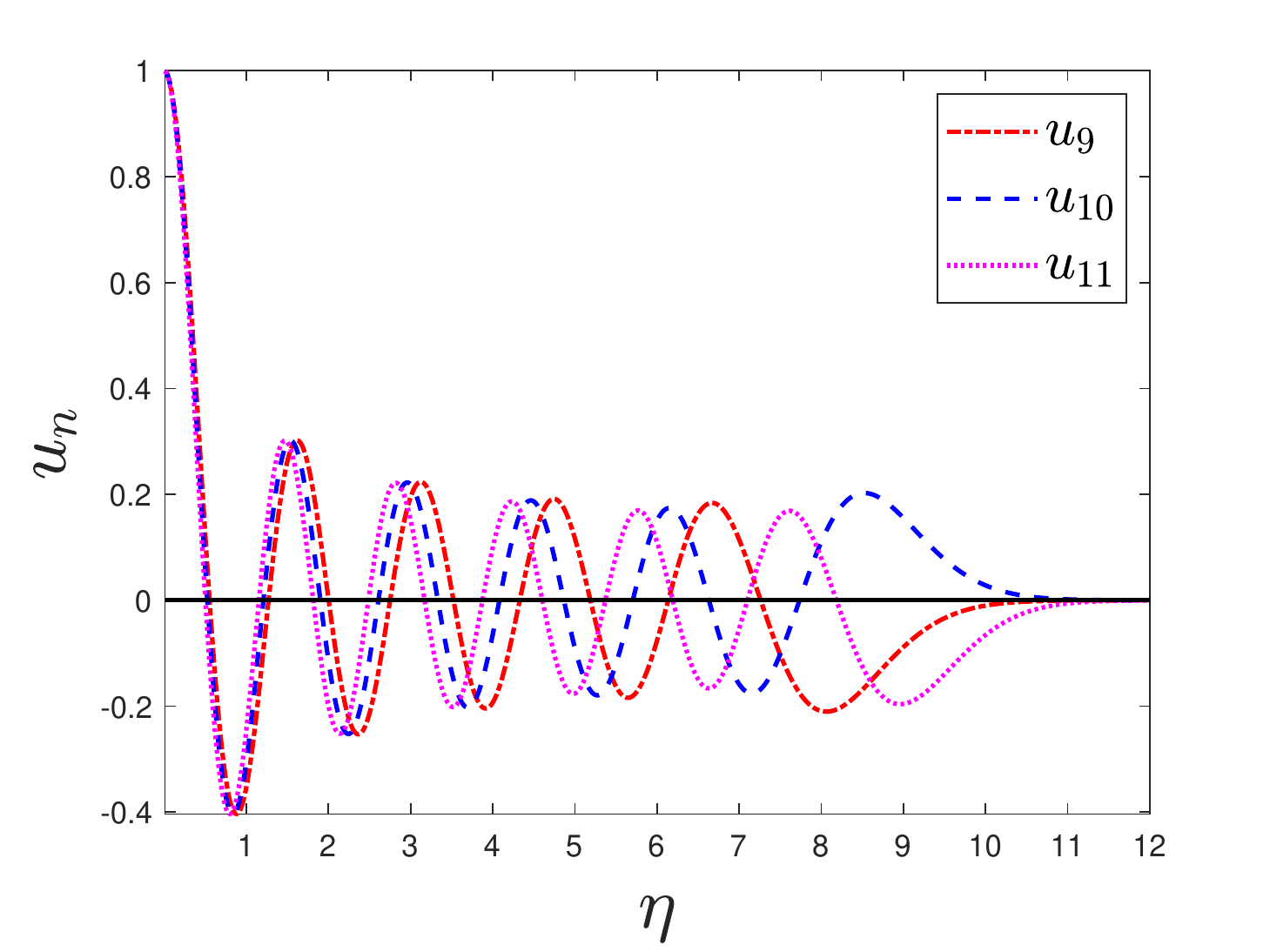}
\centering
\centering
\caption{Illustration of $u_n$ versus $\eta=R/l_B$ for $n=9,10,$ and $11$, where $R$ is the distance between two impurities and $l_B$ is the magnetic length.}\label{Fig_1}
\end{figure}

\section{Iterative Diagonalization}
In this section, we will describe how to use the numerical iterative diagonalization procedure adopted in the method of NRG to analyze the Kondo effect in strong magnetic fields. We emphasize that our calculation focuses on the low energy states at final iteration, which corresponds to the low energy states of reduced Hamiltonian $H_1$ or $H_2$. At each iteration, the iterative procedure disregards high energy states. The convergence of our results indicates the energy scale separation of high energy and low energy modes.  
\subsection{Transform diagonal matrix into hopping matrix}
The first step of the iterative diagonalization is to transform the diagonalized Hamiltonian of conduction electron into hopping Hamiltonian.
Explicitly, it means to find the hopping energy $t_m$,  atomic energy $\epsilon_m$, and the transformation $A_{n\sigma}=u_{n,m}f_{m\sigma}$ such that the kinetic energy $\sum_{n\sigma}\xi^c_{n \sigma} A^{\dag}_{n\sigma}A_{n\sigma} $ in Eq.(\ref{H002}) becomes
\begin{eqnarray} \label{NRG1} \notag
&&\sum_{n\sigma}\xi^c_{n \sigma} A^{\dag}_{n\sigma}A_{n\sigma} \\
&&=\sum_{m=1,\sigma}^{N_t} \epsilon_m f^{\dag}_{m\sigma}f_{m\sigma}+ t_m (f^{\dag}_{m\sigma}f_{m+1\sigma}+h.c.).
\end{eqnarray}
Here we have re-numbered the Landau-level index $n$ from the original range $(N_{min}, N_{max})$ to $(1,N_t)$ with the understanding that the energy of the  Landau level changes to $\varepsilon^c_n=\varepsilon_B(n-1+N_{min}+\frac{1}{2})$.
$\{ u_{n,m} \}$ is the transformation matrix transforming $A_{n\sigma}$ to $f_{n\sigma}$. As the impurities operators 
$d^\dag_{1\sigma}$ and $d^\dag_{2\sigma}$ are already put at site 1 (see Eq.(\ref{hybridization}),  $u_{n,1}$ is given by the coupling coefficient to the impurity. For the remaining components $u_{n,m}$,  we consider a vector $v_{m,n}$ for fixed $n$ as the eigenvector to the Hamiltonian with hopping $t_m$ and on-site energy $\epsilon_m$ such that  $v_{m,n}$  satisfies
\begin{eqnarray}  \label{vnm}
\begin{pmatrix} \epsilon_1& t_1 & \\\ t_1 & \epsilon_2 & t_2 \\\ & t_2 & ... \end{pmatrix}\quad
\begin{pmatrix} v_{1,n}\\v_{2,n}\\... \end{pmatrix}\quad
=
\xi_n\begin{pmatrix} v_{1,n}\\v_{2,n}\\... \end{pmatrix}\quad.
\end{eqnarray}
Here $\xi_n$ is the eigenvalue.  Clearly, the desired transformation $\{ u_{n,m} \}$ is given by $u_{n,m}=v_{m,n}$. Hence $v_{1,n}=u_{n,1}$.
From  Eq. (\ref{vnm}), it is easy to find the recurrence relation for $v_{m,n}$
\begin{eqnarray} \label{NRG2} \notag
&& t_mv_{m+1,n}=(\xi_n-\epsilon_m)v_{m,n}-t_{m-1}v_{m-1,n}, \\ \notag
&& \:\:\:\:\:\:\:\:\:\:\:\:\:\:\:\:\:\:\:\:\:\:\:\:\:\:\:\:\:\:\:\:\:\:\:\:\:\:\:\:\:\:\:\:\:\:\:\:\:\:\:\:\:\: \mbox{for} \: 1<m<N_t, \\ \notag
&& t_1v_{2,n}=(\xi_n-\epsilon_1)v_{1,n}, \\
&& t_{N_t}v_{N_t-1,n}=(\xi_n-\epsilon_{N_t})v_{N_t,n}.
\end{eqnarray}
$v_{m,n}$ (and thus $u_{n,m}$) can be found by the initial condition  $v_{1,n}=u_{n,1}$. The orthogonality requirement of transformation $\hat{v}=\left(v_{mn}\right)$ is $\hat{v}\hat{v}^T=I$, i.e.
$\sum_{n}v_{m,n}v_{m',n}=\delta_{m,m'}$.
Hence by multiplying $v_{m,n}$ in Eq.(\ref{NRG2}) and summing over $n$,  we find
\begin{eqnarray} \label{NRG3}
\epsilon_m=\sum_{n}\xi_n^2v_{m,n}^2.
\end{eqnarray}
\begin{figure}[tp] \label{Fig_2_1}
\includegraphics[height=2.5in, width=3.2in]{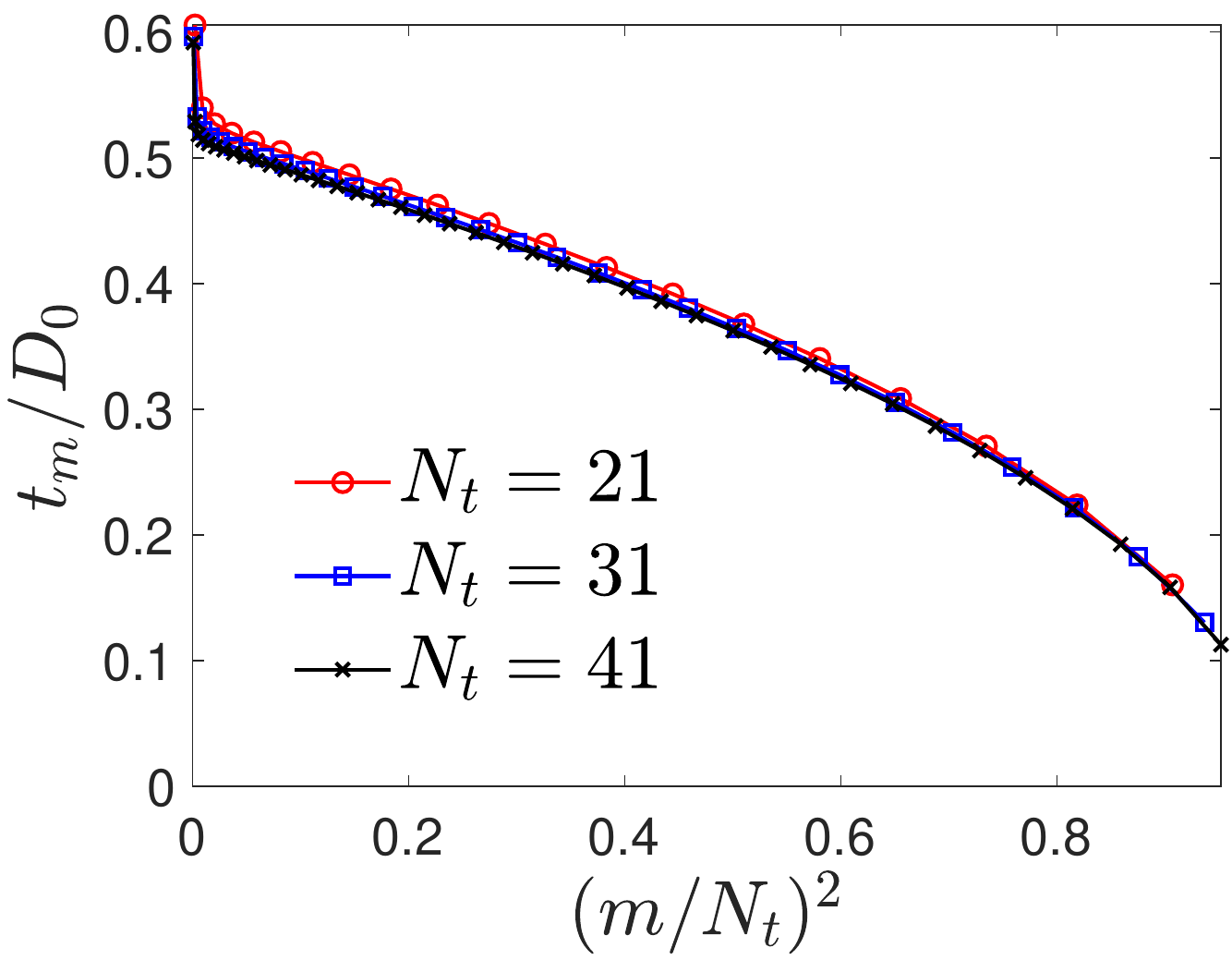}
\centering
\centering
\caption{The hopping amplitudes, $t_m$, follow the scaling form $t_m/D = f(m/N_t)$ with $f(x)$ being roughly in the form of $a-bx$..}
\end{figure}
By using Eq.s(\ref{NRG2}) and (\ref{NRG3}), $v_{1,n}$, and the normalized condition $\sum_{n}v^2_{m,n}=1$,
all $v_{m,n}$ and $t_m$ can be determined numerically. Numerically,  as shown in Fig.~2, we find that for a single impurity, $\epsilon_m=0$ and $t_m$ follows a scaling form
$t_m/D = f(m/N_t)$ with $f(x)$ being roughly in the form of $a-bx$. Furthermore,  we find that for $N_t\gtrsim 36 $, due to the accumulation of numerical error that  includes the error introduced by diagonalization in each recurrence step, the orthogonality of first and final states is poor. Hence in our calculation, we limit our calculations to systems with $N_t<30$  so that the absolute value of the inner product between first and final states $<10^{-10}$.
The error generated from this step can be neglected, as it is the order $10^{-10}$ of $\varepsilon_B$ when comparing
the eigenvalue of hopping matrix with the original diagonal matrix. Note that $t_m$ does not decay exponentially but exhibits a square-root like decay form for large $m$ as indicated in the Supplementary Information\cite{sup}. The accuracy problem related to decay form of $t_m$ will be discussed in the last subsection of this section.

\subsection{ Effective 1D chain Hamiltonians for single impurity and two impurities}

The transformation of single impurity Hamiltonian $H_1$ into an effective 1D chain Hamiltonian can be achieved by setting the new annihilation operator $f_{1\sigma}=\frac{1}{\sqrt{N_t}}\sum_n A_{n\sigma}$ with $N_t=N_{max}-N_{min}+1$ being the total number of Landau levels within the energy cutoff.
This gives $u_{n,1}=\frac{1}{\sqrt{N_t}}$. With $\xi_n=\varepsilon_B (n-1+N_{min}+\frac{1}{2})-\mu$ and $u_{n,1}$, $t_m$ and $\varepsilon_m$ can be obtained  by solving Eqs.(\ref{NRG2}) and (\ref{NRG3}). After the transformation, we obtained the single impurity 1D-chain Hamiltonian as
\begin{eqnarray} \label{HW1} \notag
&&H^W_{1}=Ud_{\uparrow}^{\dag}d_{\uparrow}d_{\downarrow}^{\dag}d_{\downarrow}+\sum_{\sigma} \xi^d_{\sigma}d_{\sigma}^{\dag}d_{\sigma} +\bar{\Gamma}\sum_{\sigma}\left(d^{\dag}_{\sigma}f_{1\sigma}+h.c.\right) \\ \notag
&&+\sum^{N_t}_{m=1,\sigma}\epsilon_{m\sigma}f_{m\sigma}^{\dag}f_{m\sigma} +\sum^{N_t-1}_{m=1,\sigma} t_m\left( f_{m\sigma}^{\dag}f_{m+1\sigma}+h.c.\right),\\
\end{eqnarray}
where the effective coupling between the impurity electrons and conduction electrons is $\bar{\Gamma}=\left(\frac{\Gamma\varepsilon_B}{\pi}\right)^{1/2}N_t^{1/2}$.
Similarly, the transformation of two-impurities Hamiltonian $H_2$ can be achieved by setting $f_{1\sigma}=\frac{1}{\sqrt{\sum_n x^2_n}}\sum_n x_nA_{n\sigma}$
and $g_{1\sigma}=\frac{1}{\sqrt{\sum_n y^2_n}}\sum_n y_nB_{n\sigma}$. The resulting two-impurities 1D-Chain Hamiltonian is given by
\begin{eqnarray} \label{HW2}\notag
&&H^W_{2}=U\sum_{j=1,2}d_{j\uparrow}^{\dag}d_{j\uparrow}d_{j\downarrow}^{\dag}d_{j\downarrow}+\sum_{j=1,2,\sigma} \xi^d_{\sigma} d_{j\sigma}^{\dag}d_{j\sigma}\\ \notag
&& +\sum^{N_t}_{m=1,\sigma}\epsilon^f_{m\sigma}f_{m\sigma}^{\dag}f_{m\sigma}+\epsilon^g_{m\sigma}g_{m\sigma}^{\dag}g_{m\sigma} \\ \notag
&& +\sum^{N_t-1}_{m=1,\sigma} \left( t^f_m f_{m\sigma}^{\dag}f_{m+1\sigma}+t^g_m g_{m\sigma}^{\dag}g_{m+1\sigma}+H.c.\right)  \\ \notag
&& +\sum_{\sigma}\bar{\Gamma}_f(d^{\dag}_{1\sigma}+d^{\dag}_{2\sigma})f_{1\sigma}+\bar{\Gamma}_g(d^{\dag}_{1\sigma}-d^{\dag}_{2\sigma}) g_{1\sigma}+h.c., \\
\end{eqnarray}
where the effective coupling  are $\bar{\Gamma}_f= (\Gamma\varepsilon_B / \pi \sum_{\{n\}}x_n^2 )^{1/2}$ and
$\bar{\Gamma}_g= (\Gamma\varepsilon_B / \pi \sum_{\{n\}}y_n^2 )^{1/2}$.

\begin{figure}[tp] \label{Fig_2}
\includegraphics[height=1.8in, width=3.2in]{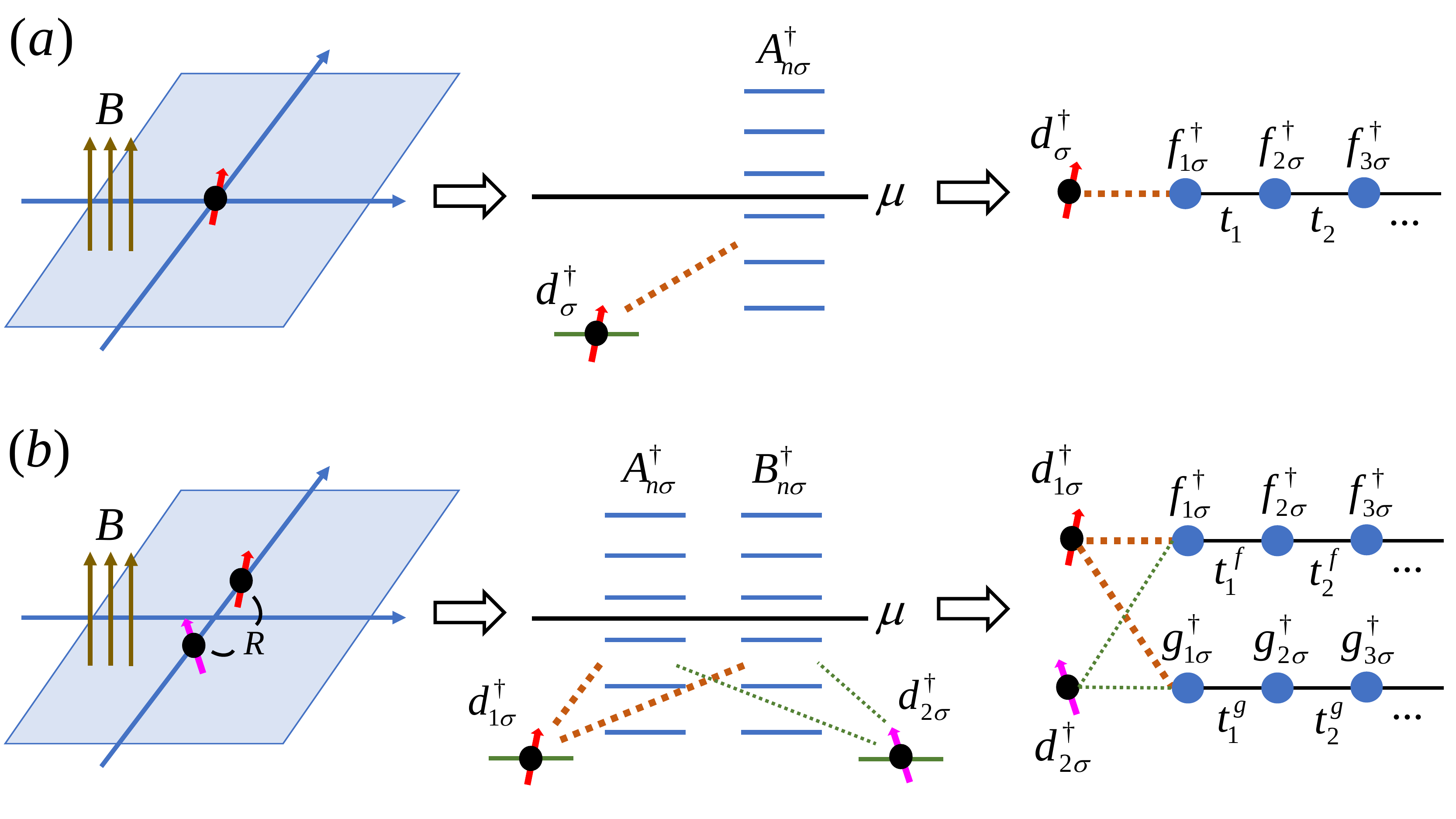}
\centering
\centering
\caption{A schematic diagram shows steps for transforming multi-impurity Anderson model to the effective 1D-chain Hamiltonian for (a) single impurity and (b) two impurities.}
\end{figure}

Finally, the procedure for transformation of the reduced Hamiltonian $H_1$ and $H_2$ to an effective 1D-Chain Hamiltonian $H_w$ is schematically illustrated in Fig.~3.

\subsection{Details of the 1D-Chain model calculation}

Details of iteratively diagonalizing single impurity one channel Hamiltonian $H^W_1$ is well known\cite{Bulla,Krishna1,Krishna1}.
We will only show details of iteratively diagonalizing two or higher channels Hamiltonian.

We start with $H^W_2$ shown in  Eq.(\ref{HW2}), and relabel $d_{1\sigma}=P_{-1\sigma}$, $d_{2\sigma}=P_{0\sigma}$,
$f_{m\sigma}=P_{2m-1\sigma}$, $g_{m\sigma}=P_{2m\sigma}$. Then the Hamiltonian includes the Hubbard terms $P^{\dag}_{-1\uparrow}P_{-1\uparrow}P^{\dag}_{-1\downarrow}P_{-1\downarrow}$,
$P^{\dag}_{0\uparrow}P_{0\uparrow}P^{\dag}_{0\downarrow}P_{0\downarrow}$, the charge terms $\sum_\sigma{P^{\dag}_{n\sigma}P_{n\sigma}}$, the Zeeman interaction terms $P^{\dag}_{0\uparrow}P_{0\uparrow}-P^{\dag}_{0\downarrow}P_{0\downarrow}$ and
several hopping terms, such as $P^{\dag}_{n\sigma}P_{n+1\sigma}+h.c.$, $P^{\dag}_{n\sigma}P_{n+2\sigma}+h.c.$ and
$P^{\dag}_{n\sigma}P_{n+3\sigma}+ h.c.$.
The Hamiltonian is the summation of these terms and commutes with total charge, $\hat{Q}_N$, and $z$-component of total spin, $\hat{S}^z_N$, which are given by
\begin{eqnarray} \label{QN1} \notag
&&\hat{Q}_N=\sum_{n}^N (P^{\dag}_{n\uparrow}P_{n\uparrow}+P^{\dag}_{n\downarrow}P_{n\downarrow}-1), \\
&&\hat{S}^z_N=\sum_{n,\sigma\sigma'}^N P^{\dag}_{n\sigma}\frac{\tau^z_{\sigma\sigma'}}{2}P_{n\sigma'}.
\end{eqnarray} Here $\tau^z$ is the z-component of the Pauli matrices. $N$ is number of $P^{\dag}_{\sigma}$ operators of any chain Hamiltonian. 
Hence the Hamiltonian is block-diagonalized by using the basis specified by quantum numbers $Q$ and $S_z$. Furthermore, each block in the Hamiltonian is specified by the quantum number $(Q,S_z)$ with the energy eigenstates for each block being represented by $|Q,S_z,r\rangle_N$, where $r=1,2,3,...$ is the ordering of the eigenstates in each block.  In addition, when two impurities have the same strength in Hubbard interaction and atomic energy, $H^W_2$ possess additional parity symmetry. In order to apply this symmetry in diagonalization, it is more convenient to re-arrange operators in $H^W_2$ by using even or odd representations of operators defined by
\begin{eqnarray} \notag
&& d_{e\sigma}=\frac{1}{\sqrt{2}}(d_{1\sigma}+d_{2\sigma}), \: d_{o \sigma}=\frac{1}{\sqrt{2}}(d_{1\sigma}-d_{2\sigma}), \\
&& c_{m,e\sigma}=f_{m\sigma},\: c_{m,o\sigma}=g_{m\sigma},
\end{eqnarray}
where the subscript $e$ labels even and $o$ labels odd parity.  $H^W_2$ then becomes
\begin{eqnarray} \label{HW2_1} \notag
&&H^W_{2}=\frac{U}{2} \left[ (d_{e\uparrow}^{\dag}d_{e\uparrow}+d_{o\uparrow}^{\dag}d_{o\uparrow})(d_{e\downarrow}^{\dag}d_{e\downarrow}+d_{o\downarrow}^{\dag}d_{o\downarrow})\right] \\ \notag
&&+\frac{U}{2}\left[ (d_{e\uparrow}^{\dag}d_{o\uparrow}+d_{o\uparrow}^{\dag}d_{e\uparrow})(d_{e\downarrow}^{\dag}d_{o\downarrow}+d_{o\downarrow}^{\dag}d_{e\downarrow})\right] \\ \notag
&& +\sum_{\sigma} \xi^d_{\sigma} (d_{e\sigma}^{\dag}d_{e\sigma}+d_{o\sigma}^{\dag}d_{o\sigma}) \\ \notag
&& +\sum^{N_t}_{m=1,\sigma}\epsilon^f_{m\sigma}c_{m,e\sigma}^{\dag}c_{m,e\sigma}+\epsilon^g_{m\sigma}c_{m,o\sigma}^{\dag}c_{m,o\sigma}  \\ \notag
&& +\sum^{N_t-1}_{m=1,\sigma} \left( t^f_m c_{m,e\sigma}^{\dag}c_{m+1,e\sigma}+t^g_m c_{m,o\sigma}^{\dag}c_{m+1,o\sigma}+H.c.\right)  \\
&& +\sum_{\sigma}\sqrt{2}\bar{\Gamma}_f d^{\dag}_{e\sigma} c_{1,e\sigma}+\sqrt{2} \bar{\Gamma}_g d^{\dag}_{o\sigma} c_{1,o\sigma}+h.c..
\end{eqnarray}
Using the relabelling notations, $d_{e\sigma}=P_{-1\sigma}$, $d_{o\sigma}=P_{0\sigma}$,
$c_{m,e\sigma}=P_{2m-1\sigma}$, $c_{m,o\sigma}=P_{2m\sigma}$, the parity operator is given by $\hat{P}_N=(-1)^{\hat{O}_N}$, where
\begin{eqnarray} \label{Parity} 
&&\hat{O}_N=\sum_{n=\mbox{even},\sigma} P^{\dag}_{n\sigma}P_{n\sigma}.
\end{eqnarray}
Note that for single impurity, one uses the quantum number $(Q,S_z)$ to perform iterative diagonalization; while for two impurities, one can either use  $(Q,S_z)$ or $(Q,S_z,P)$ for  iterative diagonalization. In the following, we will present the formalism of Hamiltonian involved in each iteration by using either the set of quantum number $(Q,S_z)$ or $(Q,S_z,P)$.
\subsubsection{Iterative diagonalization in $(Q,S_z)$ basis}
Let $|Q,S_z,r\rangle_N$ denotes  eigenstates of $H_N$ such that
\begin{eqnarray} \label{App_D2}
H_N |Q,S_z,r\rangle_N = E_N(Q,S_z,r)|Q,S_z,r\rangle_N,
\end{eqnarray}
where $r=1,2,3,...$ labels the ordering of the eigenstates in each block specified by $Q$ and $S_z$. 

When adding a new site with Fermion operator $P^{\dag}_{N+1\sigma}$, new states have to include extra particles 
(holon, one particle or two particles) at the new site so that we define new basis states as follows
\begin{eqnarray} \label{App_D3} 
&& |q,r,1\rangle_{N+1} \equiv |0;Q+1,S_z,r\rangle_{N}, \nonumber \\ 
&&|q,r,2\rangle_{N+1} \equiv |\uparrow_{N+1};Q,S_z-\frac{1}{2},r\rangle_{N}, \nonumber \\
&&|q,r,3\rangle_{N+1} \equiv |\downarrow_{N+1};Q,S_z+\frac{1}{2},r\rangle_{N}, \nonumber \\
&&|q,r,4\rangle_{N+1}\equiv  |\uparrow\downarrow_{N+1};Q-1,S_z,r\rangle_{N} , 
\end{eqnarray}
where we collectively denote the quantum number $(Q+1,S_z)$ by $q$ and the relevant states on the right hand side are defined by using $P^{\dag}_{N+1\sigma}$ as
\begin{eqnarray} \label{App_D4} \notag
&&|0;Q+1,S_z,r\rangle_{N}\equiv|Q+1,S_z,r\rangle_{N}, \\ \notag
&&|\uparrow_{N+1};Q,S_z-\frac{1}{2},r\rangle_{N}\equiv P^{\dag}_{N+1\uparrow}|Q,S_z-\frac{1}{2},r\rangle_{N},  \\ \notag
&&|\downarrow_{N+1};Q,S_z+\frac{1}{2},r\rangle_{N}\equiv P^{\dag}_{N+1\downarrow}|Q,S_z+\frac{1}{2},r\rangle_{N},  \\ \notag
&&|\uparrow\downarrow_{N+1};Q-1,S_z,r\rangle_{N}\equiv P^{\dag}_{N+1\uparrow}P^{\dag}_{N+1\downarrow}|Q-1,S_z,r\rangle_{N}.
\end{eqnarray}
Note that $|q,r,i\rangle_N$ can be also written in the form $|Q,S_z,r,i\rangle_N$, and the states $|q,r,i\rangle_N$ with $i=1,2,3$, and $4$ are built from energy eigenstates $|Q,S_z,r\rangle_{N-1}$ with number of sites being $N-1$ but they are not  energy eigenstates for number of sites being $N$. 


The Hamiltonian with an extra site is given by $H_{N+1}=H_{N}+H^I_{N,1}$, where the non-vanishing matrix elements of hopping Hamiltonian in the same block labelled by $(Q,S_z)$,
$H^I_{N,1}=\sum_{\sigma}P^{\dag}_{N\sigma}P_{N+1\sigma}+h.c.$, can be expressed as matrix elements of $P^{\dag}_{N\sigma}$ in 
the basis of energy eigenstates.
\begin{eqnarray} \label{App_D5} \notag
&& \langle q,r,1|H^I_{N,1}|q,r',2\rangle_{N+1}=\langle Q+1,S_z,r|P^{\dag}_{N\uparrow}|Q,S_z-\frac{1}{2},r'\rangle_{N}, \\ \notag
&& \langle q,r, 3|H^I_{N,1}|q,r',4\rangle_{N+1}=-\langle Q,S_z+\frac{1}{2},r|P^{\dag}_{N\uparrow}|Q-1,S_z,r'\rangle_{N}, \\ \notag
&& \langle q,r,1|H^I_{N,1}|q,r',3\rangle_{N+1}=\langle Q+1,S_z,r|P^{\dag}_{N\downarrow}|Q,S_z+\frac{1}{2},r'\rangle_{N}, \\ \notag
&& \langle q,r,2|H^I_{N,1}|q,r'4\rangle_{N+1}=\langle Q,S_z-\frac{1}{2},r|P^{\dag}_{N\downarrow}|Q-1,S_z,r'\rangle_{N}. \\ 
&& \\ \notag
\end{eqnarray}
Hence we need to calculate $\langle Q,S_z,r|P^{\dag}_{N\sigma}|Q',S_z',r'\rangle_{N}$. For this purpose, we  first note the following identities by using the definition of $|Q,S_z,r,i\rangle_N$
\begin{eqnarray} \label{App_D6} \notag
&&\langle Q+1,S_z+\frac{1}{2},r,2|P^{\dag}_{N\uparrow}|Q,S_z,r,1\rangle_{N}=1, \\ \notag
&&\langle Q+1,S_z+\frac{1}{2},r,4|P^{\dag}_{N\uparrow}|Q,S_z,r,3\rangle_{N}=1, \\ \notag
&&\langle Q+1,S_z-\frac{1}{2},r,3|P^{\dag}_{N\downarrow}|Q,S_z,r,1\rangle_{N}=1, \\
&&\langle Q+1,S_z-\frac{1}{2},r,4|P^{\dag}_{N\downarrow}|Q,S_z,r,2\rangle_{N}=-1.
\end{eqnarray}
Clearly, to get $\langle Q,S_z,r|P^{\dag}_{N\sigma}|Q',S_z',r'\rangle_{N}$, we need the transformation matrix $U_N(Q,S_z,r,w,i)$ that diagonalizes the block in $H_N$ labelled by $Q$ and $S_z$.  In other word,
$U_N(Q,S_z,r,w,i)$ connects the energy eigenstate $|Q,S_z,r\rangle_{N}$ with the basis states $|Q,S_z,w,i\rangle_{N}$ by
\begin{eqnarray} 
|Q,S_z,r\rangle_{N}=U_N(Q,S_z,r,w,i)|Q,S_z,w,i\rangle_{N}. \label{UN}
\end{eqnarray}
Here both $r$ and $w$ label the ordering of the state and $i=1-4$ with labels $1$, $2$, $3$, and $4$ representing the $N$ state for adding a holon, one spin-up particle, one spin-down particle, and two-particles to $N-1$ state as we go from $N-1$ to $N$ states.
 Using $U_N(Q,S_z,r,w,i)$ , one can compute $\langle Q,S_z,r|P^{\dag}_{N\sigma}|Q',S_z',r'\rangle_{N}$. 

As a example, consider
the computation of $\langle q,r,1|H^I_{N,1}|q,r',2\rangle_{N+1}$, which can be reduced to matrix element of $P^{\dag}_{N\uparrow}$ in
the energy eigenstates as
$\langle Q+1,S_z,r|P^{\dag}_{N\uparrow}|Q,S_z-\frac{1}{2},r'\rangle_{N}$. By using Eq.(\ref{UN}), one can express
$|Q,S_z-\frac{1}{2},r'\rangle_{N}$ in terms of $|Q,S_z-\frac{1}{2},r',w,1 \rangle_{N-1}$ or $|Q,S_z-\frac{1}{2},r',w,3 \rangle_{N-1}$.
Similarly, $|Q+1,S_z,r \rangle_{N}$ can be expressed in terms of $|Q+1,S_z,r,w,2 \rangle_{N-1}$ or $|Q+1,S_z,r,w,4 \rangle_{N-1}$. We find
\begin{eqnarray} \notag
&& \langle q,r,1|H^I_{N,1}|q,r',2\rangle_{N+1}  \\ \notag
&& =U^*_N(Q+1,S_z,r,w,2)U_N(Q,S_z-\frac{1}{2},r',w,1)\\ \notag
&& +U^*_N(Q+1,S_z,r,w,4)U_N(Q,S_z-\frac{1}{2},r',w,3).
\end{eqnarray}
Similarly, we can find all other matrix elements.

Similarly, the non-vanishing matrix elements of hopping Hamiltonian $H^I_{N,2}=\sum_{\sigma}P^{\dag}_{N\sigma}P_{N+2\sigma}+H.c.$
can be obtained as
\begin{eqnarray} \label{App_D7} \notag
&& H^{I,(q,1,2)_{N+2}}_{N,2}=\langle Q+1,S_z,r|P^{\dag}_{N\uparrow}|Q,S_z-\frac{1}{2},r'\rangle_{N+1}, \\ \notag
&& H^{I,(q,3,4)_{N+2}}_{N,2}=-\langle Q,S_z+\frac{1}{2},r|P^{\dag}_{N\uparrow}|Q-1,S_z,r'\rangle_{N+1}, \\ \notag
&& H^{I,(q,1,3)_{N+2}}_{N,2}=\langle Q+1,S_z,r|P^{\dag}_{N\downarrow}|Q,S_z+\frac{1}{2},r'\rangle_{N+1}, \\ \notag
&& H^{I,(q,2,4)_{N+2}}_{N,2}=\langle Q,S_z-\frac{1}{2},r|P^{\dag}_{N\downarrow}|Q-1,S_z,r'\rangle_{N+1}, \\
\end{eqnarray}
where $H^{I,(q,i,j)_{N+2}}_{N,2}$ is a shorthand symbol for $\langle q,r,i|H^I_{N,2}|q,r',j\rangle_{N+2}$.

Similar construction shows that the non-vanishing matrix elements of $P^{\dag}_{N\sigma}$ in basis states with $N+1$ sites are given by
\begin{eqnarray} \label{App_D8} \notag
&&\langle q,r,1|P^{\dag}_{N\sigma}|q',r',1\rangle_{N+1}=\langle Q+1,S_z,r|P^{\dag}_{N\sigma}|Q'+1,S_z',r'\rangle_{N},\\ \notag
&&\langle q,r,2|P^{\dag}_{N\sigma}|q',r',2\rangle_{N+1}=-\langle Q,S_z-\frac{1}{2},r|P^{\dag}_{N\sigma}|Q',S_z'-\frac{1}{2},r'\rangle_{N},\\ \notag
&&\langle q,r,3|P^{\dag}_{N\sigma}|q',r',3\rangle_{N+1}=-\langle Q,S_z+\frac{1}{2},r|P^{\dag}_{N\sigma}|Q',S_z'+\frac{1}{2},r'\rangle_{N},\\ \notag
&& \langle q,r,4|P^{\dag}_{N\sigma}|q',r',4\rangle_{N+1}=\langle Q-1,S_z,r|P^{\dag}_{N\sigma}|Q'-1,S_z',r'\rangle_{N}. \\
\end{eqnarray}
One can thus obtain $\langle Q,S_z,r|P^{\dag}_{N\sigma}|Q',S_z',r'\rangle_{N+1}$ by using $U_{N+1}(Q,S_z,r,w,i)$, $U_{N}(Q,S_z,r,w,i)$, Eq.(\ref{App_D6}), and Eq.(\ref{App_D8}).
Note that the above procedure can be easily generalized to the hopping Hamiltonian $\sum_{\sigma}P^{\dag}_{N\sigma}P_{N+l\sigma}+h.c.$ for arbitrary number $l$.

\subsubsection{Iterative diagonalization in $(Q,S_z,P)$ basis}

Let $|Q,S_z,P,r\rangle_N$ denots the eigenstates of $H_N$, i.e.
\begin{eqnarray} \label{App_D9}
H_N |Q,S_z,P,r\rangle_N = E_N(Q,S_z,P,r)|Q,S_z,P,r\rangle_N, \nonumber \\
\end{eqnarray}
where $P=\pm 1$ labels the parity of states.

The basis states for new states when adding new site Fermions, $P^{\dag}_{N+1\sigma}$, depend on whether $N$ is even or odd.
For $N$ is even, basis states are given by
\begin{eqnarray} \label{App_D10} \notag
&&|q,r,1\rangle_{N+1}=|0;Q+1,S_z,r\rangle_{N}, \\ \notag
&&|q,r,2\rangle_{N+1}=|\uparrow_{N+1};Q,S_z-\frac{1}{2},P,r\rangle_{N},  \\ \notag
&&|q,r,3\rangle_{N+1}=|\downarrow_{N+1};Q,S_z+\frac{1}{2},P,r\rangle_{N},  \\
&&|q,r,4\rangle_{N+1}=|\uparrow\downarrow_{N+1};Q-1,S_z,P,r\rangle_{N},
\end{eqnarray}
while for $N$ is odd, basis states are given by
\begin{eqnarray} \label{App_D11} \notag
&&|q,r,1\rangle_{N+1}=|0;Q+1,S_z,r\rangle_{N}, \\ \notag
&&|q,r,2\rangle_{N+1}=|\uparrow_{N+1};Q,S_z-\frac{1}{2},-P,r\rangle_{N},  \\ \notag
&&|q,r,3\rangle_{N+1}=|\downarrow_{N+1};Q,S_z+\frac{1}{2},-P,r\rangle_{N},  \\
&&|q,r,4\rangle_{N+1}=|\uparrow\downarrow_{N+1};Q-1,S_z,P,r\rangle_{N},
\end{eqnarray}
where $q$ is the shorthand of $(Q+1,S_z,P)$. Note that  the parity of any many-particle state changes sign 
when adding an odd number of Fermions, while the parity stays the same when adding even number of Fermions.

Note that to preserve the parity, the hopping Hamiltonian now only includes $H^I_{N,2}=\sum_{\sigma}P^{\dag}_{N\sigma}P_{N+2\sigma}+h.c.$ terms, whose non-vanishing matrix elements are given by
\begin{eqnarray} \label{App_D12} \notag
&& H^{I,(q,1,2)_{N+2}}_{N,2}\\ \notag
&&=\langle Q+1,S_z,P,r|P^{\dag}_{N\uparrow}|Q,S_z-\frac{1}{2},(-1)^{N+1} P,r'\rangle_{N+1}, \\ \notag
&& H^{I,(q,3,4)_{N+2}}_{N,2}\\ \notag
&&=-\langle Q,S_z+\frac{1}{2},(-1)^{N+1}P,r|P^{\dag}_{N\uparrow}|Q-1,S_z,P,r'\rangle_{N+1}, \\ \notag
&& H^{I,(q,1,3)_{N+2}}_{N,2}\\ \notag
&&=\langle Q+1,S_z,P,r|P^{\dag}_{N\downarrow}|Q,S_z+\frac{1}{2},(-1)^{N+1} P,r'\rangle_{N+1}, \\ \notag
&& H^{I,(q,2,4)_{N+2}}_{N,2}\\ \notag
&&=\langle Q,S_z-\frac{1}{2},(-1)^{N+1} P,r|P^{\dag}_{N\downarrow}|Q-1,S_z,P,r'\rangle_{N+1}. \\
\end{eqnarray}

The rest steps for computing the matrix elements $\langle Q,S_z,P, r|P^{\dag}_{N\sigma}|Q',S_z',P,r'\rangle_{N}$ are the same as
 what were done for basis states using $(Q,S_z)$.  Here relevant identities, similar to Eqs.( \ref{App_D6}) are given by
\begin{eqnarray} \label{App_D13} \notag
&&\langle Q+1,S_z+\frac{1}{2},P^*,w,2|P^{\dag}_{N\uparrow}|Q,S_z,P,w,1\rangle_{N}=1, \\ \notag
&&\langle Q+1,S_z+\frac{1}{2},P^*,w,4|P^{\dag}_{N\uparrow}|Q,S_z,P,w,3\rangle_{N}=1, \\ \notag
&&\langle Q+1,S_z-\frac{1}{2},P^*,w,3|P^{\dag}_{N\downarrow}|Q,S_z,P,w,1\rangle_{N}=1, \\ \notag
&&\langle Q+1,S_z-\frac{1}{2},P^*,w,4|P^{\dag}_{N\downarrow}|Q,S_z,P,w,2\rangle_{N}=-1, \\
\end{eqnarray}
where $P^*=(-1)^{N+1}P$, and relevant matrix elements are give by
\begin{eqnarray} \label{App_D14} 
&&\langle q,r,1|P^{\dag}_{N\sigma}|q',r',1\rangle_{N+1} \\ \notag
&&\:\:\:\:\:\:\:\:\:\:\:\:\:\:\:\:=\langle Q+1,S_z,P,r|P^{\dag}_{N\sigma}|Q'+1,S_z',P',r'\rangle_{N},\\ \notag
&&\langle q,r,2|P^{\dag}_{N\sigma}|q',r'2\rangle_{N+1}\\ \notag
&&\:\:\:\:\:\:\:\:\:\:\:\:\:\:\:\:=-\langle Q,S_z-\frac{1}{2},P,r|P^{\dag}_{N\sigma}|Q',S_z'-\frac{1}{2},P',r'\rangle_{N},\\ \notag
&&\langle q,r,3|P^{\dag}_{N\sigma}|q',r'3\rangle_{N+1}\\ \notag
&&\:\:\:\:\:\:\:\:\:\:\:\:\:\:\:\:=-\langle Q,S_z+\frac{1}{2},P,r|P^{\dag}_{N\sigma}|Q',S_z'+\frac{1}{2},P,'r'\rangle_{N},\\ \notag
&& \langle q,r,4|P^{\dag}_{N\sigma}|q',r', 4\rangle_{N+1}\\
&&\:\:\:\:\:\:\:\:\:\:\:\:\:\:\:\:=\langle Q-1,S_z,r|P^{\dag}_{N\sigma},P|Q'-1,S_z',P',r'\rangle_{N}. \nonumber \\
\end{eqnarray}

\subsection{Numerical Iterative Diagonalization and Error Analysis}

Based on the effective 1D-chain Hamiltonian, one can perform the iterative diagonalization procedure by diagonalizing $H^W_1$ and $H^W_2$ iteratively\cite{Bulla,Krishna1,Krishna2}. Here eigenstates of 1D-chain Hamiltonian $H^W_1$ are classified by the quantum number, charge $Q$, z-component of total spin $S_z$ and additional parity number $P$ in $H^W_2$. In the iterative diagonalization, one derives matrix elements of the effective 1D-chain Hamiltonian of $N+1$ sites (single impurity) or $N+2$ sites (two impurities) from eigenstates of $N$ sites\cite{sup}. The resulting  effective 1D-chain Hamiltonian of $N+1$ sites (single impurity) or $N+2$ sites are then exactly diagonalized. For each iteration step, numbers of eigenstates kept are  $N_{tr}=10000$ for single impurity and $N_{tr}=6000$ for the  two impurities case. By comparing with the exact excitation energies when $\Gamma=0$, the relative error of our calculations at $k_BT<<\varepsilon_B$ can be estimated to be less than $0.01\%$ for the single impurity and less than $1\%$ for two impurities.

Note that in typical research on Kondo effects, the impurity is embedded in a continuous conduction band.
The main difficulty in typical Kondo problem arises from infinite degrees of freedom for excitation energies lower than any given finite temperature $T$.
Therefore, the suitable approximated NRG Hamiltonian in wild temperature range to zero temperature limit is required. This gives the requirement of the exponentially-decayed $t_m$ by the perturbation argument in Wilson's original NRG paper.

However, in our considered situation which is primarily at zero temperature, the system is under strong magnetic field and is at temperature $T$ much lower than the Landau level energy spacing $\varepsilon_B$.  Only small number ($N_t<20$) of Landau levels is within the energy cut-off $D$. Therefore, below our interested temperature, only few degrees of freedom for excitation energies are allowed  Although $t_m$ does not decay exponentially, for small $N_t$ and large number of kept states in each iteration, the iterative diagonalization procedure still provides low energy excitations and states with high accuracy. In particular, we find that our results converges as $N_t$ increases. This indicates that the energy scale of high energy and low energy modes separates in our approach.  Furthermore, we find that $t_m$ follows a scaling form $t_m/D = f(m/N_t)$ with $f(x)$ being roughly in the form of $a-bx^2$. The existence of this scaling form implies that there is a finite-size rescaling involved when one goes from one scale to another, indicating the close relation of our method to the renormalization group analysis.

\section{Phases of Many-Body Ground state}
In this section, we will describe the emergent phases in the many-body ground state. Before we describe these phases, we shall first examine the application regime of 
our calculations.
In our iterative diagonaloization procedure, there are $N_t$ Landau levels with discrete energies within the cutoff $D$. For a given temperature $T$, there are two regimes: (1)
Regime of weak magnetic fields in which $\varepsilon_B \ll k_BT \ll D$ so that $N_t$ is essentially infinite and number of Landau levels below $k_BT$is also essentially infinite. 
This is the regime that one may apply the Wilson’s  discretization scheme. (2) Regime of strong magnetic fields in which there are finite number $N_t$ of Landau levels 
within the cutoff $D$. For typical strong magnetic fields around 10 Tesla, $N_t$ is the order of $10$ to $10^2$. This is the situation concerned in our NRG scheme.
In this regime, one needs to consider Kondo effects from finite  number of Landau levels. Furthermore, because our iterative diagonalization procedure is accurate for low energy 
excitations, it further sets a limit that the temperature is much lower than the  Landau level energy spacing $\varepsilon_B$, i.e. $k_B T <  \varepsilon_B$. From the view of renormalization group method, the system is finite and one can not perform infinite iterations and goes to the fixed point. Instead, as it is well-
known, there will be finite size effects and one needs to do finite-size scaling to get results for infinite systems. This is particularly true for scaling functions and 
scaling exponents. In this work, however, we are interested in phases of the ground states. Therefore, the finite size effect is not particularly important as one will see 
in the following that changes of $N_t$ has limited effects. 

\subsection{Transition between doublet and singlet ground state and the phase diagram when $g_c=g_d=0$}
To realize how the Kondo physics affects many-body ground state, we start with the single impurity case when $g_c=g_d=0$.
In this case, the Hamiltonian possesses additional $SU(2)$ symmetry of total spin, which allows us to classify eigenstates
by total spin $S$\cite{Krishna1,Krishna2}, which provides more accurate description of states and excitation energy.
Therefore, we shall label states by using quantum numbers $Q$ and $S$ and denote eigenstates by $|Q,2S+1,r \rangle$ and energy eigenvalues by $E(Q,2S+1,r)$.

When $\Gamma=0$, the ground state is $|0,2,r_{min} \rangle$, where the labelling $r_min$ is used to indicate that the energy of the ground state is the minimum of all $|0,2,r\rangle$.  The state  $|0,2,r_{min} \rangle$ is doubly degenerated and will persist to be an eigenstate but may not be the ground state when $\Gamma \neq 0 $. In Fig.~4(a), we show the coupling of the impurity to Landau levels schematically. Here even number of Landau quantized bands in the energy cut-off $D$ is shown. The chemical potential is set in the central two Landau levels, and below the upper level with $0.15 \varepsilon_B$. In Fig.~4(b), we show the energy difference $\Delta E$
between first few low energy states and doublet states $|0,2,r_{min} \rangle$ at different $\rho J$. Here in the calculations, we change the parameter $\Gamma$ but in the plot shown in Fig.~4(b), we use $\rho J$ as the variable for the x-axis. $\rho J$ is related to $\Gamma$ by\cite{Hewson}
\begin{eqnarray} \label{rho_J}
\rho J=\frac{\Gamma}{\pi}(\frac{1}{|U+\xi_d|}+\frac{1}{|\xi_d|}).
\end{eqnarray}
For each given $\Gamma$, from numerical calculation, one obtains low energy states $|Q,2S+1,r \rangle$ and spectrum $E(Q,2S+1,r)$ classified by the quantum number $(Q,2S+1)$.
To make the competition between two ground states more clear, the energy difference $\Delta E$ is taken as $E(Q,2S+1,r)-E(0,2,r_{min})$ as shown in Fig.~4.
\begin{figure}[hbtp] \label{Fig3}
\includegraphics[height=1.5in, width=3.3in]{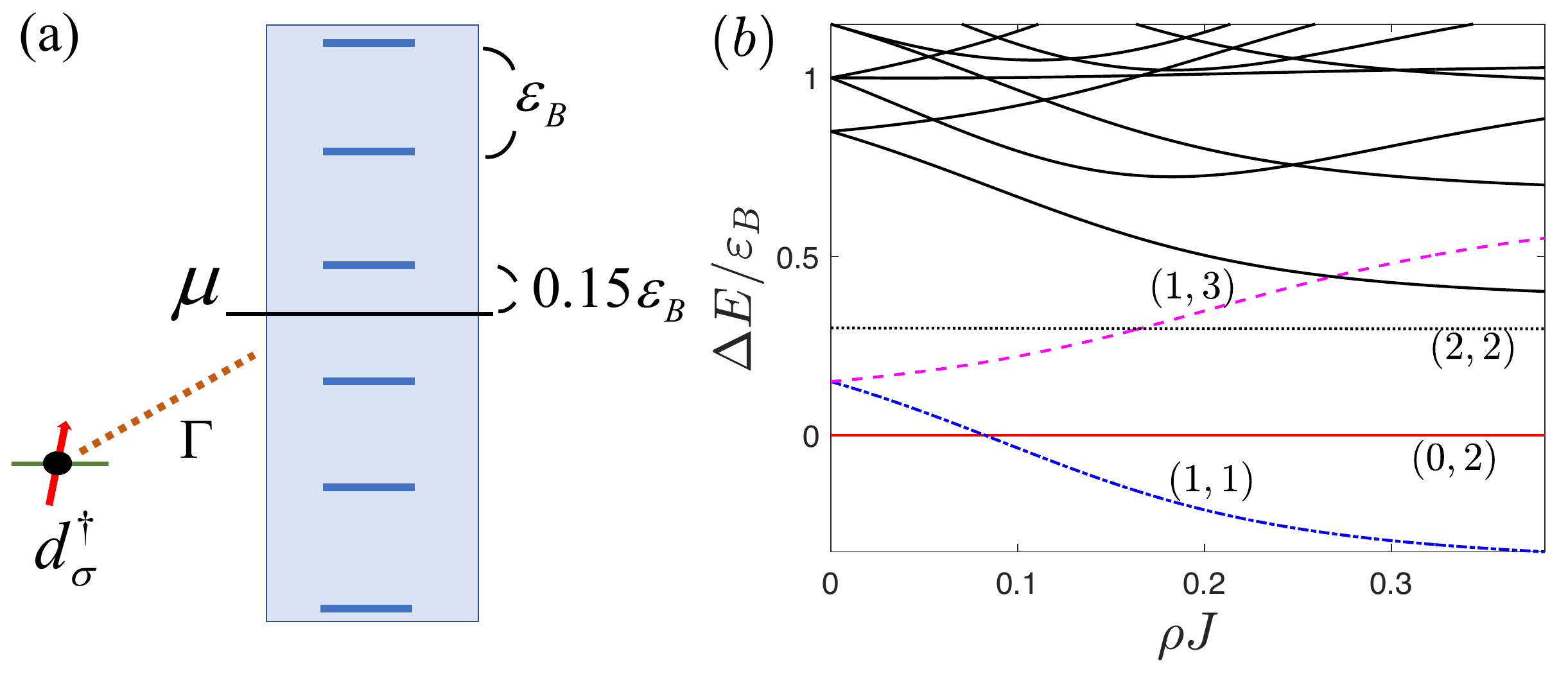}
\centering
\centering
\caption{ (a) Schematic diagram for single impurity coupling to Landau levels within the cut-off $D$ (shaded area). (b) Many-particle eigen-energy difference, $E(Q,2S+1,r)-E(0,2,r_{min})$, obtained from the iterative diagonalization method, where numbers shown in brackets labels $(Q,2S+1)$ and  $\rho J=\frac{\Gamma}{\pi}(\frac{1}{|U+\xi_d|}+\frac{1}{|\xi_d|})$.}
\end{figure}

\begin{figure}[hbtp] \label{Fig4}
\includegraphics[height=2.95in, width=3.3in]{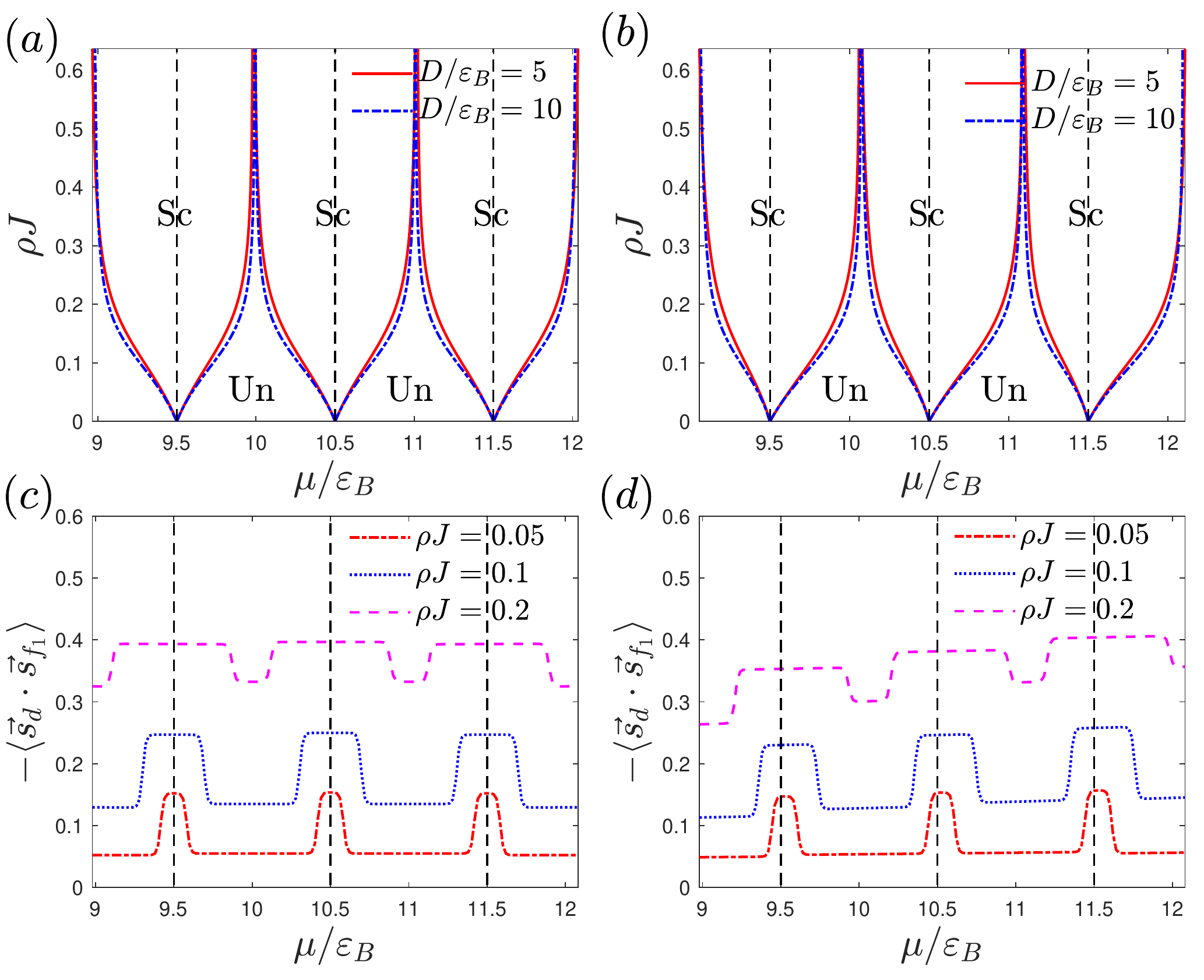}
\centering
\centering
\caption{ Phase diagram of many-body ground states in the absence of Zeeman splitting. Here states are classified by degeneracy $2S+1$. For $2S+1=1$ phase is denoted as $Sc$(Screened), for 
$2S+1=2$ phase is denoted as $Un$(Unscreened). Here $\rho J=\frac{\Gamma}{\pi}(\frac{1}{|U+\xi_d|}+\frac{1}{|\xi_d|})$, and $\xi_d=-5D$.
The vertical black dashed line mark the position of Landau level before hybridization.
(a) $U=10D$, the impurity Hamiltonian possesses particle-hole symmetry. (b)$U=10^4 D$, this is the typical infinite $U$ region, which can be described by the  t-J model. (c) $U=10D$,  quantum oscillation of spin-spin correlation strength $\langle \vec{s}_d \cdot \vec{s}_{f_1} \rangle$, which is negatively enhanced when the system enters into the screened phase. (d)$U=10^4 D$.}
\end{figure}

Let us first examine low energy many-body states at $\Gamma=0$ as shown in Fig.~4(b).
When $\Gamma=0$, the ground state $|E_0(\Gamma=0)\rangle$ of the system is the direct product of the ground state for the impurity ground state $|E^d_0(\Gamma=0)\rangle$  and the ground state of conduction electrons  $|E^c_0(\Gamma=0)\rangle$. Here
the ground state of the conduction electrons $|E^c_0(\Gamma=0)\rangle$ is the state with all levels below $\mu$ being doubly occupied.
For $\xi^d<0$ and $U+\xi^d>0$, the impurity prefers singly occupied and hence the ground state of the spin is doubly degenerated due to spin. As a result, the ground state $|E_0(\Gamma=0)\rangle$ possesses $Q=0$ and degeneracy $2S+1=2$.
The first excited state $|E_1(\Gamma=0)\rangle$ has $4$ degeneracies. This corresponds to the
addition of a charge into the ground state of the conduction electrons (so its $Q=1$) with energy $0.15 \varepsilon_B$,
and the degeneracy  $4$  comes from the spin of impurity and the spin of added charge.
The second excited state $|E_2(\Gamma=0)\rangle$ has $2$ degeneracies, which correspond to the
addition of two charges into the ground state of the conduction electrons with total energy $0.3 \varepsilon_B$.
The degeneracy comes from the spin of the impurity.

When $\Gamma$ is turned on, the Kondo spin-spin interaction starts to show up between the impurity and conduction electron.
However, many-particle states of conduction electrons must carry spins so that it can screen the spin of the impurity.
Therefore, as shown in Fig.~4(b), the competition between  single charge excitation energy ($0.15\varepsilon_B$) at $\Gamma=0$ and the Kondo interaction energy when $\Gamma$ is finite,  results in the quantum phase transition between the local moment doublet and Kondo singlet state, as displayed in the crossover between red solid line and blue dashed dotted line.
In addition, spin-triplet states are shown in pink dashed line where its energy increases when $\Gamma$ increases as predicted by the sign of Kondo interaction.
Furthermore, we notice that the energy difference between $|E_0(\Gamma=0)\rangle$ and $|E_2(\Gamma=0)\rangle$ is almost unchanged  when $\Gamma$ increases to large values, which also agrees with the argument that the Kondo interaction only significantly affects many-particles of conduction electrons.

Based on the degeneracy of ground state at different $\mu$ and $\Gamma$, we plot the phase diagram in the parameter space of $\mu$ and $\rho J$ in Figs.~5 (a) and 5(b), where for doublet phase ($2S+1=2$) we denote it as $Un$(Unscreened) phase and for singlet phase ($2S+1=1$) we denote it as $Sc$(Screened) phase. The Kondo screening feature in these phases can be checked by examining the spin-spin correlation $\langle \vec{s}_d \cdot \vec{s}_{f_1} \rangle$ between the impurity $d$ and the first site $f_1$ of the 1D-chain. This
is shown in Fig.~5(c) and 5 (d), in which we see that in agree with the Kondo screening feature, the spin-spin correlation is negatively enhanced when the system enters into the screened phase.

\subsection{Phase diagram and magnetic moment when $g_c \neq 0 $ and $g_d\neq0$}

When $g_c \neq 0 $ and $g_d\neq0$, the Zeeman splitting term in Hamiltonian generally breaks the typical temperature-driven Kondo effect in weak fields\cite{ImpMag2}.
To simplify the numerical calculation, we define new $\tilde{g}$-factors as 
\begin{eqnarray} \label{g-fac}
\tilde{g}_c=g_c\frac{\mu_B B }{\varepsilon_B}, \tilde{g}_d=g_d\frac{\mu_B B }{\varepsilon_B}.
\end{eqnarray}
Note that when the effective electron mass $m^*_e$ is equal to the free-electron mass, one has $\tilde{g}_c=g_c$ and $\tilde{g}_d=g_d$.

Since $g_c \neq 0$ and $g_d\neq0$, the total spin $S$ is not a good quantum number.
Therefore, phase diagram at zero temperature is obtained by keeping track of change of the quantum numbers $S_z$ in the lowest energy state. From $S_z$ (the $z$-component of total spin) of the lowest energy state at $\rho J=0$ and finite $\rho J$, one obtains change of $S_z$ that is due to the Kondo interaction as 
\begin{equation} 
\Delta S_z\equiv S^0_z(\rho J)-S^0_z(\rho J=0),
\end{equation}
where the superscript $0$ indicates that $S^0_z$ is $S_z$ of the ground state.  Using $\Delta S_z$, we identify phases of the system as shown in Figs.~6(a), 6(b), 7(a), and 7(b). Note that the system is composed by electrons and hence possible values of total spin are half-integers, i.e., $S=0,1/2,1,3/2,....$. Hence  possible values of $\Delta S_z$ are $0$, $\pm1/2$, $\pm 1$,...

When $\tilde{g}_d>0$, $\tilde{g}_c>0$, and $\rho J=0$, $S^{imp}_z$ of the impurity in ground state is $-1/2$; while  $S^c_z$ of conduction electrons can be $-1/2$ or $0$, depending on whether the chemical potential lies between two Zeeman-split Landau levels  or not. Therefore, total $S_z$ at $\rho J=0$ is equal to $-1$ or $-1/2$, i.e., $S^0_z(\rho J=0)=-1$ or $-1/2$. 
When $\rho J>0$, $S^{imp}_z$ can be screened or unscreened.  Clearly, if $S^{imp}_z$ is unscreened,  we have $\Delta S_z=0$. This is the situation when $S^{imp}_z=-1/2$ and $S^c_z=0$ as conduction electrons have no spin to screen the impurity.
In general, finite $\langle S^{imp}_z \rangle<0$ of the impurity in together with the Kondo interaction generates an effective negative g-factor $J\langle S^{imp}_z\rangle S^c_z$. This changes $S^c_z$ of conduction electrons from $-1/2$ to $0$ or $1/2$, which corresponds to $\Delta S_z = 1/2$ or  $\Delta S_z = 1$ respectively, which is consistent with $\Delta S_z$ anticipated for half-integer systems. Thus by exploring $\Delta S_z$, we obtain unscreened phases with $\Delta S_z = 0$, and the Kondo screening state with $\Delta S_z=1/2$ and $\Delta S_z=1$ labelled by $Sc$ and $Sc^*$ respectively. In the $Sc^*$ phase, both the impurity spin and spin of conduction electrons vanishes due to Kondo screening. Clearly, as shown in Figs. 6(a), 6(b), 7 (a), and 7(b), phases of the ground state oscillate among "Sc", "Un", and "Sc$^*$" states as the chemical potential $\mu$ changes. 

\begin{figure}[hbtp] \label{Fig5}
\includegraphics[height=1.6in, width=3.3in]{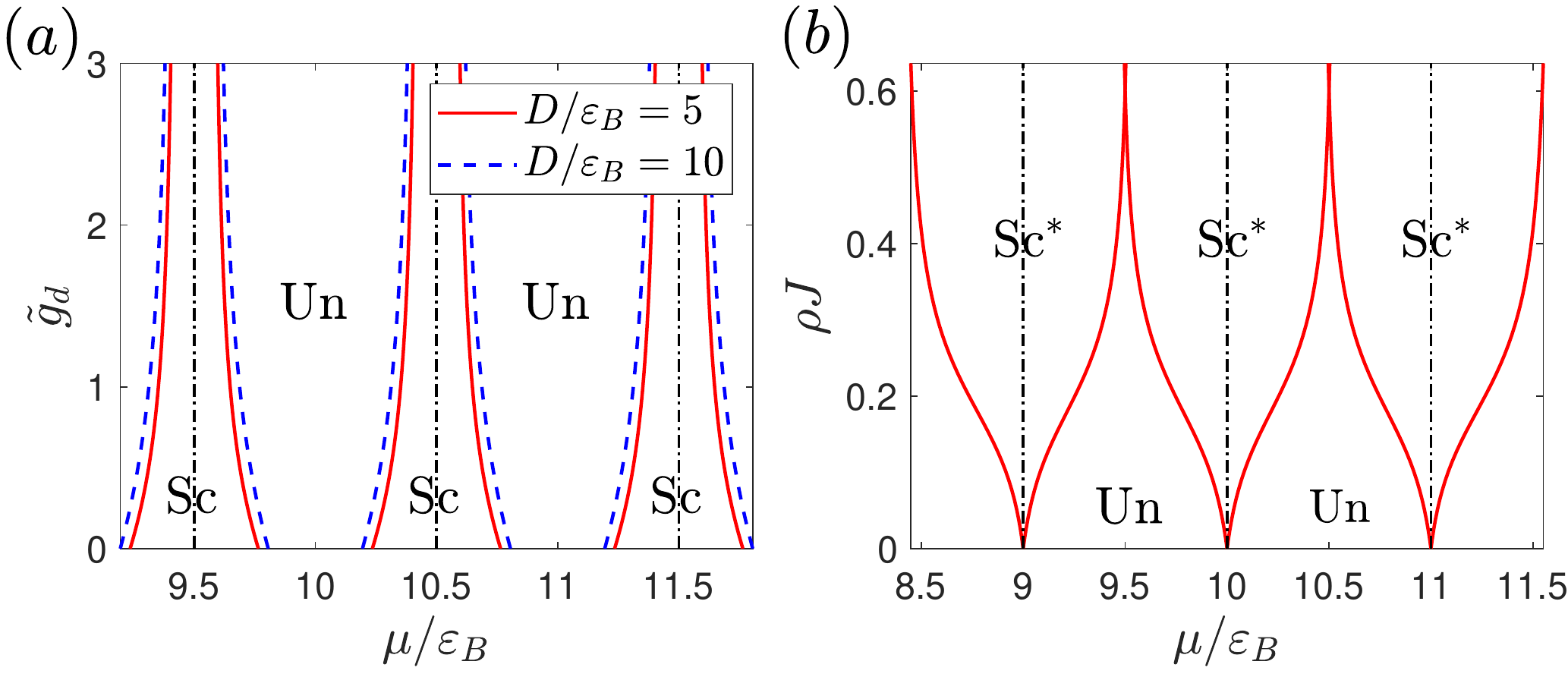}
\centering
\centering
\caption{ Phase diagram of many-body ground states in the presence of Zeeman splitting. Here states are classified by $\Delta S_z\equiv S^0_z(\rho J)-S^0_z(\rho J=0)$ . 
"Sc" denotes the Kondo screening state with $\Delta S_z=1/2$, "Sc*" denote the Kondo screening state with $\Delta S_z=1$, and "Un" denotes the unscreened impurity state.The parameter is taken as $\rho J=\frac{\Gamma}{\pi}(\frac{1}{|U+\xi_d|}+\frac{1}{|\xi_d|})$, $U=10D$, $\xi_d=-5D$.
The vertical black dashed line mark the position of Landau level before hybridization.
(a) $\tilde{g}_c=0$, $\rho J=0.127$. (b) $\tilde{g}_c=\tilde{g}_d=2$, $D/\varepsilon_B=10$.}
\end{figure}

\begin{figure}[hbtp] \label{Fig6}
\includegraphics[height=3.1in, width=3.4in]{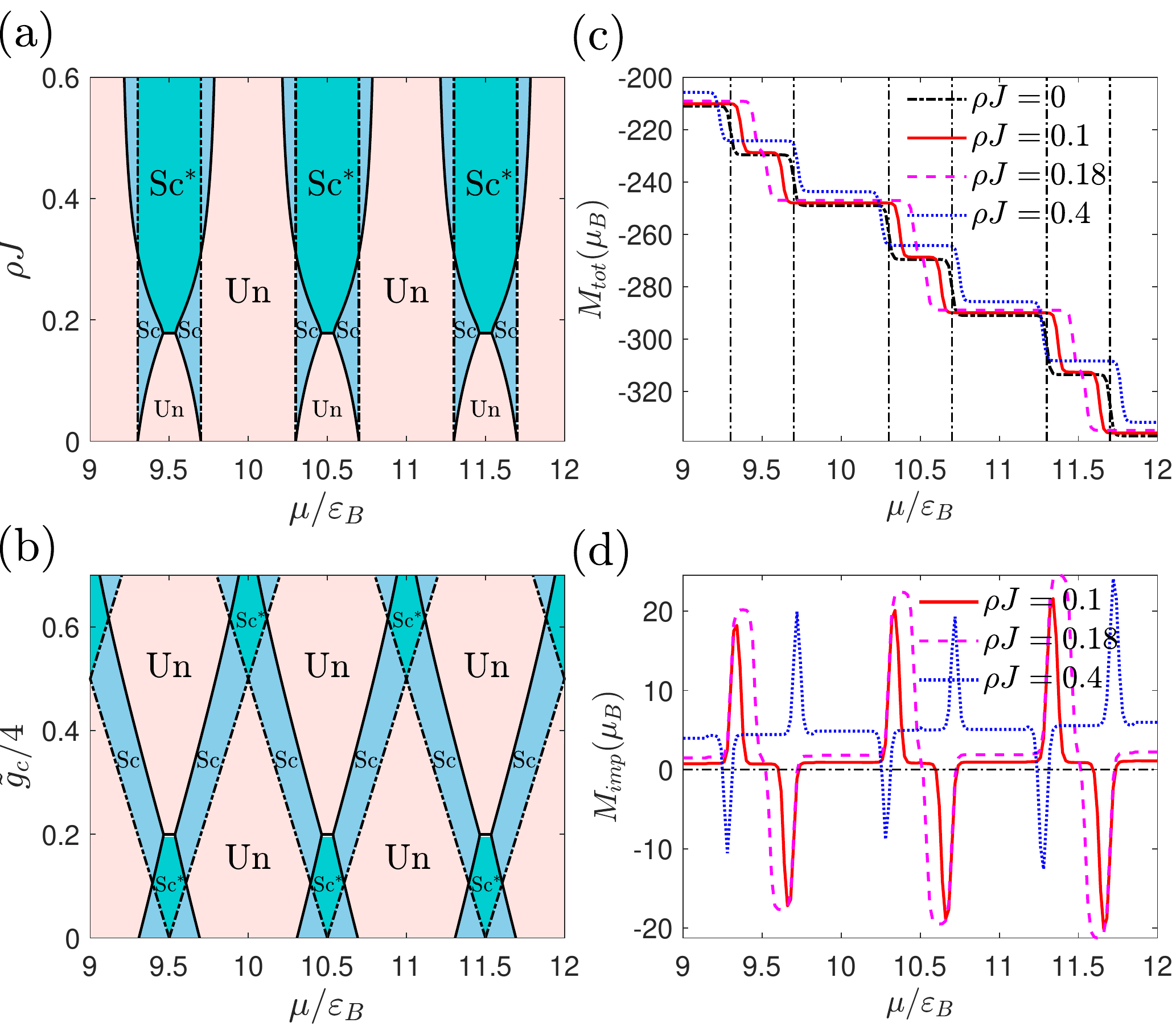}
\centering
\centering
\caption{  (a) Phase diagram of single Kondo impurity in the parameter space, $\rho J$ versus chemical potential $\mu$. Here $g_c=0.8$, $g_d=2$, "Sc" denotes the Kondo screening state with $\Delta S_z=1/2$, "Sc*" denote the Kondo screening state with $\Delta S_z=1$   , and "Un" denotes the unscreened impurity state. (b)  Phase diagram of single Kondo impurity in the parameter space, $g_c$ versus chemical potential. Here $g_d$ is fixed at 2. (c) Quantum oscillation in total magnetic moment at temperature $k_B T/D=0.005$ (sum of magnetic moments of the Kondo impurity and the conduction electrons, see text for definition), the vertical black dashed line mark the position of split Landau level before hybridization is turned on. (d)  Quantum oscillation of the total magnetic moment induced by impurity. Here parameters are $\varepsilon_B/D=0.2$, $U/D=10$, $\xi_d/D=-5$, and $\rho J=\frac{\Gamma}{\pi}(\frac{1}{|U+\xi_d|}+\frac{1}{|\xi_d|})$.}
\end{figure}

This results in quantum oscillations in magnetic moments as shown in Figs.~7(c) and 7(d). Here the total magnetic moment $M_{tot}$ is computed by the definition $M_{tot}=-\frac{\partial \Omega}{\partial B}$, where $\Omega=-\frac{1}{\beta}\mbox{Tr}(e^{-\beta H})$\cite{Moment} and can be decomposed as the summation of the magnetic momentums due to the orbit moment of conduction electrons
$M_{c}$ ,the hybridization part $M_{hyb}$,  the spin moment of conduction electrons $M_{c,s}$, the spin moment of the impurity $M_{d,s}$. The detail of each parts is as follows,
\begin{eqnarray} \label{Moment1}\notag
&& M_{c}=-2\mu_B\sum_{\{n\},\sigma} (n+\frac{1}{2}) \langle A_{n\sigma}^{\dag} A_{n\sigma} \rangle, \\ \notag
&& M_{hyb}=-\mu_B (\Gamma / \pi \varepsilon_B)^{1/2}\sum_{\{n\},\sigma} \langle d_{\sigma}^{\dag}A_{n\sigma}+H.c. \rangle, \\ \notag
&& M_{c,s}=-\frac{\mu_B g_c }{2}\sum_{\{n\}} \langle A_{n\uparrow}^{\dag} A_{n\uparrow}-A_{n\downarrow}^{\dag} A_{n\downarrow} \rangle, \\
&& M_{d,s}=-\frac{\mu_B g_d }{2}\langle d_{\uparrow}^{\dag} d_{\uparrow}-d_{\downarrow}^{\dag} d_{\downarrow} \rangle.
\end{eqnarray}
Note that we only keep one channel in the Landau level that couples to the impurity in $H_1$ and the remaining $N_L-1$ channels are disregarded (cf.  Eq. (\ref{H001})).
Hence for dilute impurity systems with impurity number being $N_{imp}$, the moment of total system is given by
$ (N_L-N_{imp})(M^{0}_c+M^{0}_{c,s})+N_{imp}M_{tot} =N_L(M^{0}_c+M^{0}_{c,s})+N_{imp}M^{0}_{d,s}  +N_{imp}(M_{tot}-M^{0}_{tot})$,
where $M^{0}_c$ and $M^{0}_{c,s}$ are magnetic moments due to the orbit and spin of conduction electrons in the absence of hybridization, $\Gamma=0$.
The total magnetic moment induced by impurity, shown in Fig.~7(d), is then given by $M^{imp}_{ind}=M_{tot}-M^{0}_{tot}$, where the magnetic moment with the superscript $0$ denotes the same moment when $\Gamma=0$.
The total moment of Kondo system with dilute impurities in strong fields is equal to
\begin{eqnarray} \label{Moment2}
 M=M_0 + N_{imp} M_{imp} +  N_{imp} M^{imp}_{ind},
\end{eqnarray}
where $M_0$ is the moment of pure Landau quantized system, $M_{imp}$ is the single impurity moment equals to $g_d\mu_B$ at low temperature,
and $M^{imp}_{ind}$ is the induced moment by impurity.

\begin{figure}[t] \label{Fig7}
\includegraphics[height=2.8in, width=3.5in]{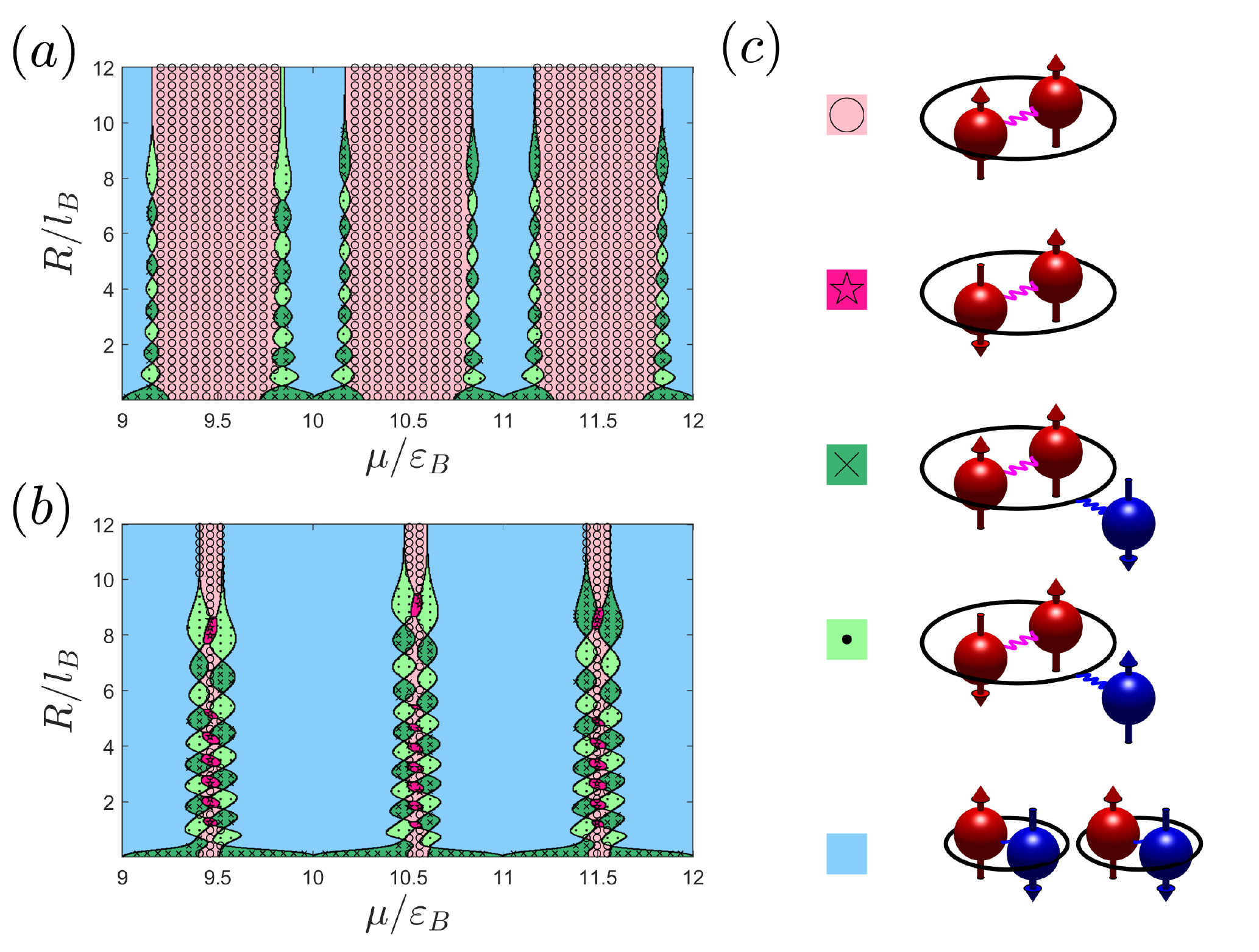}
\centering
\centering
\caption{Phase diagram of  two Kondo impurities in strong magnetic fields for (a) $\rho J =0.18$ and (b) $\rho J =0.45$. (c) different screening scenarios of two Kondo impurities in phases shown in (a) and (b). Here parameters used are $g_c=g_d=2$, $\varepsilon_B/D=1/3$, $U/D=10$, and $\xi_d/D=-5$. As determined by RKKY interaction, spins of two impurities marked by red color may form singlet state, triplet state or two independent spin-$1/2$ state. The impurity spins then get screened by conduction electrons marked by blue color.}
\end{figure}
\begin{figure}[tp] \label{Fig8}
\includegraphics[height=3.2in, width=3.6in]{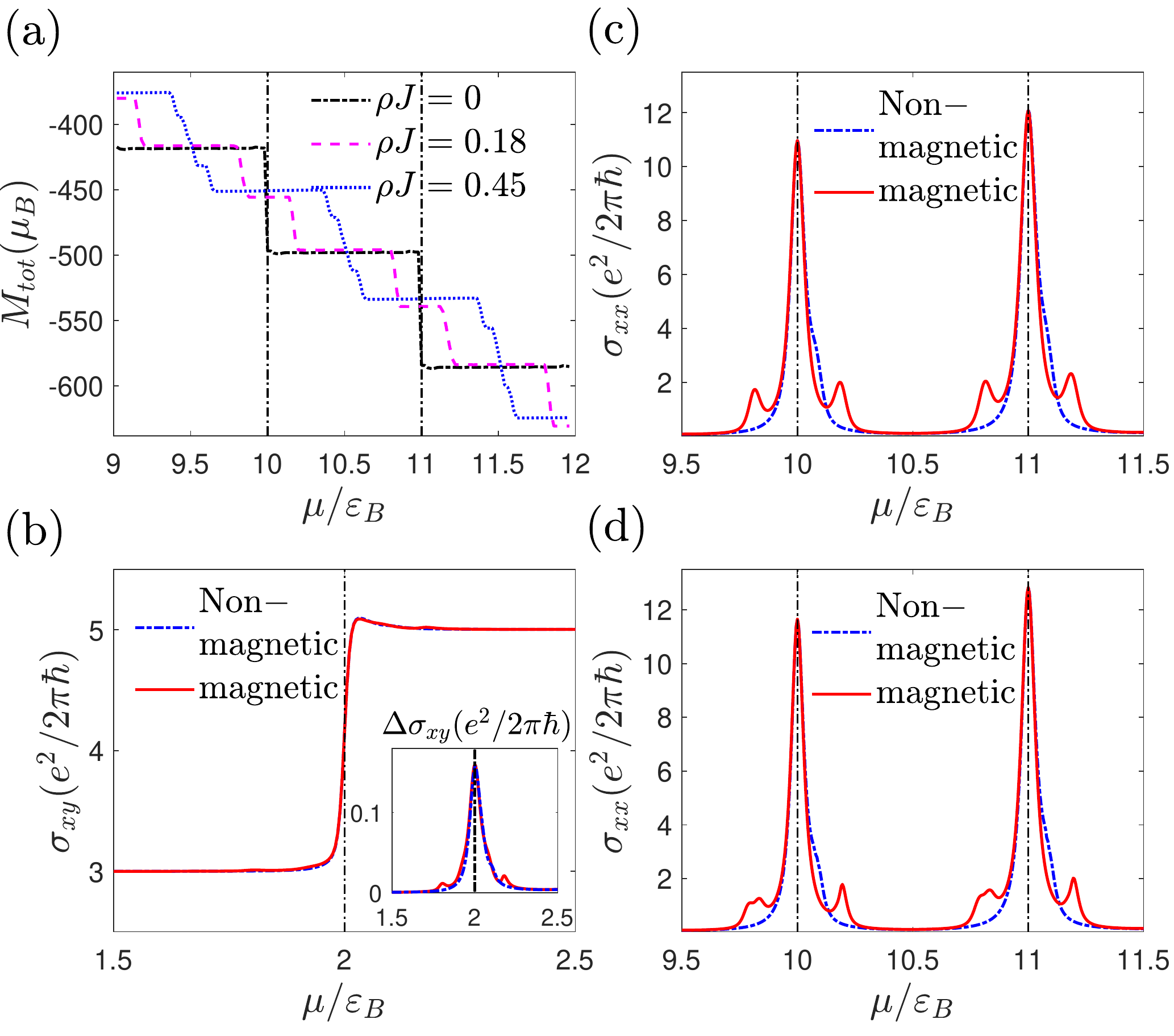}
\centering
\caption{ Quantum oscillation exhibited in two Kondo impurities under strong magnetic field. (a) Total magnetization versus the chemical potential ($\mu$) in unit of $\varepsilon_B$. Comparison of contribution of non-magnetic and magnetic impurities to Hall conductivity (b) and  longitudinal conductivity (c) for one impurity.  (d) Same comparison of two impurities
for longitudinal conductivity. Here side peaks are due to extra density of states is released from total screened state to partial screened or unscreened states (see text).  Parameters used are: $\varepsilon_B/D=1/3$, $U/D=10$, $\xi_d/D=-5$, $\rho J=0.18$, $B = 10(Tesla)$, $k_B T/D=0.005$, $R/l_B=2 16nm$, and the effective impurity scattering constant is $w_{imp}=0.0025$ (see \cite{sup})}.
\end{figure}

We now explore two Kondo impurities in the presence of strong magnetic fields. In Fig.~8, we show results of numerical iterative diagonalization on allowed phases of the system for different density of states. For two impurities in strong magnetic fields, we find that the interplay between the Kondo screening effect, RKKY interaction and quantum oscillations due to Landau levels determines the ground state of the system. Specifically, the combination of these factors results in different screening scenarios for different phases as shown in Fig.~8(c). Here as determined by RKKY interaction, spins of two impurities marked by red color may form singlet state, triplet state or two independent spin-$1/2$ state. The impurity spins then get screened by conduction electrons marked by blue color.  As a result, as shown in Fig.~8(c), we find that ground states can be unscreened triplet state, unscreened singlet state, partial screened triplet state, partial screened singlet state, and two screened spin 1/2 state. These states are characterized by different quantum number $S_z$'s that reflect the scenario such that the corresponding $S_z=1$, $0$, $1/2$, $1/2$, and $0$. Furthermore, states shown in Fig.~8(c) form different phases so that the ground state of the system with Kondo impurities oscillate between these states when either the magnetic field or the distance between Kondo impurities change  as shown in Figs.~8 (a) and (b). This oscillation leads to quantum oscillation in magnetization and conductivities as shown in Fig.~9. Here the longitudinal  conductivity $\sigma_{xx}$ and Hall conductivity $\sigma_{xy}$ are computed by
\begin{eqnarray} \label{sigma1}\notag
&& \sigma^{xx}=\sum_{s}\sigma^{xx}_{s}, \\ \notag
&& \sigma^{xx}_{s}=\frac{e^2\varepsilon_B^2}{2\pi^2\hbar}\int dE \:\frac{-\partial f(E)}{\partial E} \times\\
&&  \:\:\:\:\:\sum_{n}(n+1) \rm{Im}G^c_{n,s}(E+i\delta)\mbox{Im}G^c_{n+1,s}(E+i\delta),\\
&& \Delta\sigma^{xy}=\sum_{s} -2\frac{\rm{Im}\Sigma_s(0+i\delta)}{\varepsilon_B}\sigma^{xx}_{s}.
\end{eqnarray}
Here $s$ labels the spin of conduction electrons, $f(E)=1/(1+e^{E/k_BT})$ is the Fermi-Dirac function\cite{Ando}, and
$G^c_{n,s}(E)=\langle G^c_{n,k_y,s}(E) \rangle$ is the renormalized Green's function  for conduction electrons with the average over
all Landau degeneracies $k_y$ of $n_{th}$ Landau level being taken. In terms of the self-energy $\Sigma_{n,k_y,s}$,
$ G^c_{n,k_y,s}(E)$ can be expressed as
\begin{eqnarray} \label{Green}
G^c_{n,k_y,s}(E) = \frac{1}{E-(n+1/2+s)\varepsilon_B-\Sigma_{n,k_y,s}}.
\end{eqnarray}
Here the contribution to  the self-energy comes from the scattering of the conduction electrons by magnetic impurities. At low impurity density, this self-energy $\Sigma_{n,k_y,s}$ can be expanded in terms of the impurity density $O(n_{imp})$.
For single impurity case, $\Sigma_{n,k_y,s}=n_{imp}\Gamma/(\pi\rho) G_s^d$, while for two impurities case,
$\Sigma_{n,k_y,s}=n_{imp}\Gamma/(\pi\rho) (G_{11,s}^d+G_{22,s}^d)$.
Here $G_s^d$ generally represents the Green's function for electrons of the impurities, i.e., the $d$ electrons. 
The one-impurity Green's function is represented by $G_s^d$ and two-impurities Green's function is represented by $G_{ij,s}^d$ with $i$ and$j$ being the position of impurities. In the Lehmann representation, using eigenstates $|\alpha \rangle$ and eigen-energies $E_{\alpha} $ obtained from numerical calculations, these Green's functions are given by
\begin{eqnarray} \label{Green_d}
&&G^d_{s}=\frac{1}{Z}\sum_{\alpha \alpha'}|\langle  \alpha | d_s |\alpha' \rangle|^2\frac{e^{-\beta E_{\alpha}}+e^{-\beta E_{\alpha'}}}{E-(E_{\alpha'}-E_{\alpha})+i\delta}, \\ \notag
&&G^d_{ij,s}=\frac{1}{Z}\sum_{\alpha \alpha'}\langle \alpha | d_{is} |\alpha' \rangle \langle \alpha' | d^{\dag}_{js} |\alpha \rangle \frac{e^{-\beta E_{\alpha}}+e^{-\beta E_{\alpha'}}}{E-(E_{\alpha'}-E_{\alpha})+i\delta}.
\end{eqnarray}

Clearly, addition peak structures are seen in $\sigma_{xx}$ and $\Delta \sigma_{xy}$ shown in Fig.~9. The main peak right at the Landau level is due to the resonant scattering of conduction electrons in phase with screened impurities, while side peaks are located at the phase boundaries when extra density of states is released from total screened state to partial screened or unscreened states.  For instance, side peaks in Fig.~9(c) is due to the density of state released from screened state to unscreened state for one Kondo impurity. These peaks are important experimental signatures for phases of Kondo impurities in strong magnetic fields.

In summary, we have generalized the iterative diagonalization procedure adopted in NRG  to investigate Kondo impurities screened by discrete Landau levels. We find that the ground state generally oscillates in Kondo screened state, partially-screened, and unscreened spin states. This leads to quantum oscillations observed in magnetization and conductivity of the system.  In particular, we find  peak structures in longitudinal conductivity that reflects changes of Kondo screening phases and are important features to be observed in experiments. While we have been focusing on  one and two Kondo impurities, our results are applicable to systems with finite density of Kondo impurities.  Our results thus provide a complete characterization of phases for Kondo effect in strong magnetic fields.


\begin{acknowledgments}
This work was supported by National Science and Technology Council (NSTC), Taiwan. 
We also acknowledge support from Center for Quantum Technology within
the framework of the Higher Education Sprout Project
by the Ministry of Education (MOE) in Taiwan.
\end{acknowledgments}


\begin{thebibliography}{99}
\bibitem{Hewson} A. C. Hewson, {\it The Kondo Problem to Heavy Fermions}, Cambridge University Press, Cambridge, England, 1993.
\bibitem{mou1} Po-Hao Chou, Liang-Jun Zhai, Chung-Hou Chung, Chung-Yu Mou, and Ting-Kuo Lee, Phys. Rev. Lett. {\bf 116}, 177002 (2016)
\bibitem{Doniach} S. Doniach, in {/it Valence Instabilities and Related Narrow
Band Phenomena}, edited by R.D. Parks (Plenum, New York, 1977), p. 169; Physica B+C {\bf 91B}, 231 (1977).
\bibitem{Coqblin} J.R. Iglesias, C. Lacroix, and B. Coqblin, Phys. Rev. B {\bf 56}, 11 820 (1997).
\bibitem{twoimpurity1} C. Jayaprakash, H. R. Krishna-murthy, and J. W.  Wilkins, Phys. Rev. Lett. {\bf 47}, 737 (1981).
\bibitem{twoimpurity2} B. A. Jones and C. M.  Varma, Phys. Rev. Lett. {\bf 58}, 843 (1987).
\bibitem{twoimpurity3} B. A. Jones, C. M. Varma, and J. W. Wilkins, Phys. Rev. Lett. {\bf 61}, 125 (1988).
\bibitem{twoimpurity4} B. A. Jones and C. M. Varma, Phys. Rev. B {\bf 40}, 324 (1989).
\bibitem{twoimpurity5} J. B.  Silva, W. L. C. Lima, W. C. Oliveira, J. L. N. Mello, L. N. Oliveira, and J. W. Wilkins, Phys. Rev. Lett. {\bf 76}, 275 (1996).
\bibitem{twoimpurity6} P. Simon, R. Lopez, and Y. Oreg, Phys. Rev. Lett. {\bf 94}, 086602(2005).
\bibitem{twoimpurity7} T.  Jabben, N. Grewe, and S.  Schmitt, Phys. Rev.  B{\bf 85}, 045133 (2012).
\bibitem{ImpMag1} A. Spinelli, M. Gerrits, R. Toskovic, B. Bryant, M. Ternes, and A. F. Otte, Nat. Commun. {\bf 6}, 10046 (2015).
\bibitem{Li} G. Li, Z. Xiang, F. Yu, T. Asaba, B. Lawson,  P. Cai, C. Tinsman, A. Berkley, S. Wolgast, Y. S. Eo, Dae-Jeong Kim, C. Kurdak, J. W. Allen, K. Sun, X. H. Chen, Y. Y. Wang, Z. Fisk, and Lu Li, Science {\bf 346}, 1208 (2014).
\bibitem{Sebastian} B. S. Tan, Y.-T. Hsu, B. Zeng, M. Ciomaga Hatnean, N. Harrison, Z. Zhu, M. Hartstein, M. Kiourlappou, A. Srivastava, M. D. Johannes, T. P. Murphy, J.-H. Park, L. Balicas, G. G. Lonzarich, G. Balakrishnan, and Suchitra E. Sebastian, Sience  {\bf 349}, 287 (2015).
\bibitem{Sebastian2} H. Liu, M. Hartstein, G. J. Wallace, A. J Davies, M. C. Hatnean, M. D Johannes, N. Shitsevalova, G. Balakrishnan, and S. E Sebastian,  J. Phys.: Condens. Matter {\bf 30}, 16LT01 (2018).
\bibitem{mou2} Yen-Wen Lu, Po-Hao Chou, Chung-Hou Chung, Ting-Kuo Lee, and Chung-Yu Mou Phys. Rev. B {\bf 101}, 115102 (2020).
\bibitem{ImpGap1} C. C. Yu and M. Guerrero, Phys. Rev. B {\bf 54}, 8556 (1996).
\bibitem{ImpGap2} K. Chen and C. Jayaprakash, Phys. Rev. B {\bf 57}, 5225 (1998).
\bibitem{ImpGap3} M. R. Galpin and D. E. Logan, Phys. Rev. B {\bf 77}, 195108 (2008).
\bibitem{LLSlaveB} B. D\'{o}ra, P. Thalmeier, Phys. Rev. B {\bf 76}, 115435 (2007).



\bibitem{RKKY1} M. A. Ruderman and C. Kittel, Phys. Rev. 96, 99 (1954).
\bibitem{RKKY2} T. Kasuya, Prog. Theor. Phys. 16, 45 (1956).
\bibitem{RKKY3} K. Yosida, Phys. Rev. 106, 893 (1957).




\bibitem{Bulla} R. Bulla, T. A. Costi, and T. Pruschke, Rev. Mod. Phys. {\bf 80}, 395 (2008).



\bibitem{LLRKKY} J. Cao, H. A. Fertig, and S. Zhang, Phys. Rev. B {\bf 99}, 205430 (2019).

\bibitem{sup} See supplementary information for details of the calculation.


\bibitem{Krishna1} H. R. Krishna-murthy, J. W. Wilkins, and K. G. Wilson, Phys. Rev. B {\bf 21}, 1003 (1980).
\bibitem{Krishna2} H. R. Krishna-murthy, J. W. Wilkins, and K. G. Wilson, Phys. Rev. B {\bf 21}, 1044 (1980).
\bibitem{ImpMag2} T. A. Costi, Phys. Rev. Lett. {\bf 85}, 1504 (2000).
\bibitem{Moment} J. Knolle, and N. R. Cooper, Phys. Rev. Lett. 115, 146401 (2015).


\bibitem{Ando} T. Ando, Y. Matsumoto, and Y. Uemura, J. Phys. Soc. Jpn. {\bf 39}, 279 (1975).
\end{thebibliography}
\end{document}